\documentclass[12pt,a4paper]{article}

\usepackage{setspace,graphics}
\usepackage[dvips]{epsfig} 
\usepackage{a4,amssymb,epsfig,array,cite}
\usepackage[hypertex]{hyperref}

\newcounter{multieqs}

\newcommand{\be}{\begin{equation}}
\newcommand{\ee}{\end{equation}}
\newcommand{\eq}[1]{(\ref{#1})}

\def\nn{\nonumber}
\def\bea{\begin{eqnarray}}
\def\eea{\end{eqnarray}}

\def\beqa{\begin{eqnarray}} 
\def\eeqa{\end{eqnarray}} 
\def\beq{\begin{equation}} 
\def\eeq{\end{equation}}

\def\Tr{{\rm Tr}}

\def\a{\alpha}          
\def\b{\beta}           
  \def\C{\Gamma}  
\def\d{\delta}

 \def\L{\Lambda}

\def\cA{{\cal A}}  \def\cC{{\cal C}}
 \def\cE{{\cal E}} 
 \def\cH{{\cal H}} 
  
\def\cM{{\cal M}} \def\cN{{\cal N}}

\def\R{{\mathbb R}}
\def\C{{\mathbb C}}

\def\one{\mbox{1 \kern-.59em {\rm l}}}

\def\mmu{\mathfrak{u}}

\def\bit{\begin{itemize}}
\def\eit{\end{itemize}}

\def\({\left(}
\def\){\right)}

\def\d{\delta}

\def\uno{\mbox{1 \kern-.59em {\rm l}}}

\def\bcomment#1{}

\renewcommand{\title}[1]{\vspace{10mm}\noindent{\Large{\bf #1}}\vspace{8mm}}
\newcommand{\authors}[1]{\noindent{\large #1}\vspace{5mm}}
\newcommand{\address}[1]{{\itshape #1\vspace{2mm}}}

\begin{document}

\begin{titlepage}

\begin{flushright}
UWTHPh-2008-01\\
\end{flushright}

\begin{center}
  
\title{Emergent Gravity, Matrix Models 
 and UV/IR Mixing}

 \authors{Harald {\sc Grosse}${}^{1}$,
Harold {\sc Steinacker}${}^{2}$, Michael {\sc Wohlgenannt}${}^{3}$}

 \address{ Fakult\"at f\"ur Physik, Universit\"at Wien\\
 Boltzmanngasse 5, A-1090 Wien, Austria}

 \address{Erwin Schr\"odinger International Institute for Mathematical
 Physics\\
 Boltzmanngasse 9, A-1090 Wien, Austria}

\footnotetext[1]{harald.grosse@univie.ac.at}
\footnotetext[2]{harold.steinacker@univie.ac.at}
\footnotetext[3]{michael.wohlgenannt@univie.ac.at}

\vskip 2cm

\textbf{Abstract}

\vskip 3mm 

\begin{minipage}{14cm}%

We verify explicitly that UV/IR mixing for 
noncommutative gauge theory can be understood 
in terms of an induced gravity action, as
predicted by the identification 
\cite{Steinacker:2007dq} of gravity within
matrix models of NC gauge theory.
More precisely, we obtain the Einstein-Hilbert action
by integrating out a scalar field in the adjoint.
It arises from the well-known UV/IR mixing of NC gauge theory, 
which is carefully re-analyzed and 
interpreted in terms of gravity. The matrix model therefore
contains gravity as an IR effect, due to UV/IR mixing.

\end{minipage}


\end{center}

\end{titlepage}

\setcounter{page}0
\thispagestyle{empty}

\begin{spacing}{.3}
{
\noindent\rule\textwidth{.1pt}            
   \tableofcontents
\vspace{.6cm}
\noindent\rule\textwidth{.1pt}
}
\end{spacing}


\section{Introduction}

The idea that gravity should be related to quantum fluctuations of
space-time at the Planck scale is very old. Recently, 
a specific and concise realization of this idea has been proposed under the name of
``emergent (noncommutative) gravity''. The basic observation is that 
noncommutative (NC) gauge theory, defined through matrix models, 
contains a specific version of gravity as an intrinsic part, and 
provides a dynamical theory of noncommutative spaces.
Such a connection between gravity and NC gauge theory was 
first observed  in 
\cite{Rivelles:2002ez}, and pushed further in 
\cite{Yang:2006hj} from a somewhat different point of view; 
see also \cite{Muthukumar:2004wj} for subsequent work.
A concise form of this idea was then given in
\cite{Steinacker:2007dq} using the framework of matrix models. 
The essential point is a new, geometrical interpretation of 
the $U(1)$ sector of the standard matrix model action
for NC gauge theory.
This provides a specific form for the effective metric $G^{ab}$
in terms of a dynamical Poisson structure $\theta^{ab}$, 
which completely 
absorbs the ``would-be $U(1)$'' gauge fields of NC gauge theory. 
The correct gravitational coupling of the  
nonabelian gauge fields was also established in \cite{Steinacker:2007dq}.

One of the particularly exciting aspects of emergent NC gravity
is that it provides a simple 
prescription for the quantization of gravity, being realized as
NC gauge theory resp. Matrix Model.
In particular, it 
was pointed out in \cite{Steinacker:2007dq} that the Einstein-Hilbert 
action will be induced upon quantization, 
and that it should amount to the notorious UV/IR 
mixing in noncommutative gauge theory. This prediction is supported by
the fact that both gravity and UV/IR mixing occur only in the $U(1)$
sector of NC gauge theory. It should also explain 
the strange IR behavior \cite{Matusis:2000jf} of 
the ``would-be photons'': they are not photons but gravitons defining
a non-trivial geometric background. 
More precisely, the ``would-be $U(1)$ gauge fields'' are 
re-interpreted in terms of geometry and 
absorbed in the effective metric. 
This metric then couples to all other fields, which explains why 
the $U(1)$ sector of NC gauge theory cannot be disentangled from the 
$SU(n)$ sector. 

In this paper, we elaborate and verify this explanation 
of UV/IR mixing in
terms of gravity. This provides not only a nontrivial
consistency check for emergent NC gravity, it also paves the way
towards its quantization. We will perform a 
one-loop quantization of a scalar field coupled to the matrix model
of NC gauge theory resp. gravity in two different ways. 
In the geometrical point of view,
 we interpret the action as scalar field coupled to gravity,
which leads using standard arguments to an induced
Einstein-Hilbert action.
Second, we use the more conventional interpretation of the 
same matrix
model in terms of NC gauge theory, where integrating out the scalar
field leads to an effective action for the NC gauge fields involving
the well-known UV/IR mixing terms \cite{Minwalla:1999px}.
These two computations should agree at least in 
the IR regime, where the 
geometrical picture is expected to make sense.
We then show in detail that the Einstein-Hilbert action indeed
coincides with the effective action for the gauge fields, using the 
relation between gauge fields and the 
metric given in \cite{Steinacker:2007dq}.
This holds in the IR regime assuming a suitable effective cutoff
$\L < \L_{NC}$, where it completely captures the 
UV/IR mixing. In fact we 
need to carefully re-analyze the UV/IR mixing terms in this regime,
which has not been done in the literature so far. 

As a result, we obtain not only a non-trivial check for the basic
mechanism of emergent gravity, but also an understanding of UV/IR 
mixing in NC gauge theory. The latter has been the main obstacle 
for the physical application of NC gauge theory, because the 
physical behavior of the trace-$U(1)$ sector
forbids an interpretation as a photon. Thus the present 
point of view opens the way towards the physical application of 
NC gauge theory resp. the matrix model, and moreover
suggests a new approach towards 
the quantization and unification of gravity and gauge theory. 
In particular, the effective cutoff is related with the 
gravitational constant, rather than requiring renormalizability
in the traditional sense.
This could be realized naturally in a SUSY extension of the model
under consideration.

It is interesting to compare our explanation of UV/UR mixing 
in gauge theory with previous work in the context of string theory,
where UV/IR mixing on a brane with $B$-field background
was related to the exchange of closed string modes in the 
bulk \cite{Armoni:2001uw,Sarkar:2005jw}.
While there are some parallels in the sense that gravity 
modes are involved, our explanation is certainly simpler and 
works within the 4-dimensional framework, without additional
string modes in some higher-dimensional bulk.
Nevertheless, it might be helpful to 
understand better the relation between these different points
of view. Evidence for 4-dimensional gravitons
in a quite similar context as ours has been found previously
in \cite{Kitazawa:2005ih}, which is also related to UV/IR mixing.

The paper is organized as follows. 
We start in section \ref{sec:metric}
with a recollection of the basic mechanism how geometry and
gravity emerges from a matrix model of NC gauge theory. 
Only $U(1)$ is considered for simplicity.
Integrating out a scalar field 
leads to an induced gravity action as explained in section 
\ref{sec:ind-gravity}. We then reconsider the same 
model from the point of view of NC gauge theory
in section \ref{sec:flat-expansion}. 
The geometrical quantities and the 
induced Einstein-Hilbert action
are then expressed in terms of $U(1)$ gauge fields.
In section \ref{sec:UV-IR-mixing}, we perform the 
quantization from the gauge theory point of view,
carefully re-analyzing the 
effective action and UV/IR mixing to $O(A^2)$.
We find indeed complete agreement with the geometrical 
point of view in a suitable IR regime.  
Correction terms to the Einstein-Hilbert action are found
upon extending this IR regime.

\section{Matrix models and effective geometry}
\label{sec:metric}

Consider the matrix model with action 
\be
S_{YM} = - Tr [Y^a,Y^b] [Y^{a'},Y^{b'}] g_{a a'} g_{b b'},
\label{YM-action-1}
\ee
for
\be
g_{a a'} = \delta_{a a'} \quad \mbox{or}\quad g_{a a'} = \eta_{a a'} 
\label{background-metric}
\ee
in the Euclidean  resp.  Minkowski case. While some mathematical
aspects of this paper apply mainly to the Euclidean case, we keep the notation
general so that the Minkowski case is covered as well at least formally.
The "covariant coordinates"  $Y^a,  \,\, a=1,2,3,4$ 
are hermitian matrices, or equivalently
operators acting on a separable Hilbert space $\cH$.
We will denote the commutator of 2 matrices as 
\be
[Y^a,Y^b] = i \theta^{ab}
\label{theta-def}
\ee
so that $\theta^{ab} \in L(\cH)$ is an antihermitian\footnote{in
contrast to the conventions in \cite{Steinacker:2007dq}} 
operator-valued matrix, which is 
{\em not} necessarily a multiple of $\one_\cH$.
We focus here on configurations $Y^a$ (which need not be solutions of 
the equation of motion) which can be interpreted  via \eq{theta-def}
as quantizations 
of a Poisson manifold $(\cM,\theta^{ab}(y))$ with general
Poisson structure $\theta^{ab}(y)$. This defines the geometrical 
background under consideration, and conversely 
essentially any (local) Poisson manifold provides a possible
background $Y^a$ \cite{Kontsevich:1997vb}.
More formally, this means that there is an isomorphism of vector spaces
\be 
\cC(\cM) \to \cA \subset L(\cH)\, 
\label{map} 
\ee 
where $\cC(\cM)$ denotes some space of functions on $\cM$, 
and $\cA$ is interpreted as quantized algebra of 
functions\footnote{Roughly speaking $\cA$ is the algebra 
generated by $Y^a$, but technically one usually considers some 
subalgebra corresponding to bounded functions.} 
 on $\cM$. 
The map \eq{map} can be used to define a star product on $\cC(\cM)$.
Furthermore, we can then write
\be
\left[f,g\right] \sim i \{f(y),g(y)\} 
\ee
for $f,g \in \cA$,
where $\sim$ denotes the leading term in a semi-classical 
expansion in $\theta^{ab}$,
and $\{f,g\}$ the Poisson bracket defined by $\theta^{ab}(y)$.
$Y^a$ can be interpreted as
quantization of a classical coordinate function 
$y^a$ on $\cM$. More importantly, $Y^a$ defines a derivation on 
$\cA$ via
\be
[Y^a,f] \sim i \theta^{ab}(y) \partial_b f(y) , \qquad f \in \cA .
\label{derivation}
\ee
In this paper, we  restrict ourselves to the ``irreducible'' case,
i.e. we assume that the centralizer of $\cA$ in $\cal{H}$ is trivial.
Then any reasonable matrix (``function'') in $L(\cH)$  
can be well approximated by a function of $Y^a$.
From the gauge theory point of view in section \ref{sec:flat-expansion}, it
means that we restrict ourselves to the $\mmu(1)$ case;
this is the case of interest here since the UV/IR mixing happens in the
trace-$\mmu(1)$ sector. 
For the general case see \cite{Steinacker:2007dq}.

In order to derive the effective metric on $\cM$,
let us now consider a scalar field 
coupled to the matrix model \eq{YM-action-1}. The only possibility 
to write down kinetic terms for matter fields is through commutators 
$[Y^a,\Phi] \sim i\theta^{ab}(y) \frac{\partial}{\partial y^b} \Phi$
using \eq{derivation}. Thus consider the action
$S = S_{YM} + S[\Phi]$ where
\bea
S[\Phi] &=& - (2\pi)^2\,Tr\, \frac 12\, g_{aa'} [Y^a,\Phi][Y^{a'},\Phi]  \nn\\
&\sim& \int d^4 y\, \rho(y)\, \frac 12\, 
G^{ab}(y)\,\frac{\partial}{\partial y^a}\Phi(y) 
\frac{\partial}{\partial y^b} \Phi(y) .
\label{scalar-action-0}
\eea
Here $\sim$ indicates  the leading contribution in a semi-classical 
expansion in powers of $\theta^{ab}$, and
\be
G^{ab}(y) = \theta^{ac}(y) \theta^{b d}(y)\, g_{cd} \, 
\label{effective-metric}
\ee
 is the effective metric on $\cM$ in $y$ coordinates, 
$d s^2 = G_{ab}(y)\, dy^a dy^b$.
It plays indeed the role of a gravitational metric, because it
enters in the kinetic term for any matter coupled to the
matrix model (up to certain density factors). 
This result also holds for nonabelian gauge fields 
as shown in \cite{Steinacker:2007dq} and for fermions \cite{klammer}.
The density factor
\be
\rho(y) = |\det G_{ab}(y)|^{1/4} = (\det\theta^{ab}(y))^{-1/2} 
\equiv \L_{NC}^4(y)
\ee
is the symplectic measure on $(\cM,\theta^{ab}(y))$, which
can be interpreted as ``local'' non-commutative scale $\L_{NC}$.
We will assume in this paper that $\theta^{ab}(y)$ is nondegenerate.
Notice that the action \eq{scalar-action-0} is
invariant under Weyl rescaling of $\theta^{ab}(y)$ resp. $G^{ab}(y)$.
We can therefore write the action as
\be
S[\Phi] = \int d^4 y\, \frac 12\, \tilde G^{ab}(y)\,
 \partial_{y^a}\Phi(y) \partial_{y^b}\Phi(y) 
= \int d^4 y\, \sqrt{|\tilde G_{ab}|}\,\, \frac 12\, \Phi(y)\Delta_{\tilde G} \Phi(y) 
\label{scalar-action-geom}
\ee
where $\Delta_{\tilde G}$
is the Laplacian for the unimodular metric 
\bea
\tilde G^{ab}(y) &=& (\det G_{ab})^{1/4}\, G^{ab}(y) = \L_{NC}^4(y)\,
G^{ab}(y) \, ,\nn\\
\det \tilde G^{ab} &=& 1\, .
\label{metric-unimod}
\eea
We will often use $\tilde G^{ab}$; it is important to remember
that it is unimodular only in these $y$ coordinates.

Therefore the Poisson manifold $\cM$ naturally 
acquires a metric structure $(\cM,\theta^{ab}(y),G^{ab}(y))$, 
which is determined by the Poisson structure and the 
constant background metric $g_{ab}$ as above.
Note also that $\theta^{ac}(y)$ 
can be interpreted as a preferred frame 
or vielbein\footnote{While there are parallels with ideas in 
\cite{Madore:2000aq}, the specific mechanism 
and the geometry here is different.}, 
which is however gauge-fixed and does not admit the usual 
local Lorentz resp. orthogonal transformations. This means 
that we consider a restricted class of metrics and 
associated coordinates, where the role of 
the diffeomorphism group is replaced by the 
symplectomorphisms respecting $\theta^{ab}(y)$. For a related
discussion see \cite{Yang:2006hj}.

A linearized version of \eq{metric-unimod}  
was obtained using a similar reasoning in \cite{Rivelles:2002ez}, and the full
Seiberg-Witten expansion was given in \cite{Banerjee:2004rs} 
for the case of scalar fields. However the universal role 
(up to density factors resp. conformal rescaling) 
of \eq{effective-metric} resp. \eq{metric-unimod} 
was only recognized in \cite{Steinacker:2007dq}.
Note that this metric is {\em not} the pull-back 
of $g^{ab}$ using the change of coordinates \eq{cov-coord-1},
and it is indeed curved in general\footnote{This is in contrast to the
metrics $h_{ab}$ considered in the context of the DBI action 
\cite{Yang:2006hj} which are flat; this will be discussed in 
section \ref{sec:trafos}.}. 
It is also easy to see that in 4 dimensions, 
one cannot obtain the most general 
geometry from metrics of the form \eq{effective-metric}.
However, one does obtain a class of metrics
which is sufficient to describe the propagating
(``on-shell'') degrees of freedom of gravity, 
as well as the Newtonian limit for an 
arbitrary mass distribution.  This is discussed in 
\cite{Steinacker:2007dq} and will not be repeated here. 
We only point out that the metrics come in 
a special gauge, which is sufficient of course. The 2 propagating
degrees of freedom (helicities) of 
gravitational waves are recovered from the 2 propagating
helicities of $\mmu(1)$ gauge fields, taking advantage of the 
Poisson tensor $\theta^{ab}(y)$ \eq{graviton-1}.

\paragraph{Equations of motion.}

So far we considered arbitrary background configurations 
$Y^a$ as long
as they admit a geometric interpretation.
The equations of motion derived from the action \eq{YM-action-1} 
\be
[Y^a,[Y^{a'},Y^b]]\, g_{a a'} = 0 \, 
\label{eom}
\ee
select on-shell geometries among all possible backgrounds, such as 
the Moyal-Weyl quantum plane \eq{Moyal-Weyl}. 
In the present geometric form they amount to Ricci-flat
spaces $R_{ab}[\tilde G] \sim 0$ \cite{Rivelles:2002ez} 
at least in the linearized case. 
However since we are interested in the quantization
here, we have to consider general off-shell configurations below.

\section{Quantization and induced gravity}
\label{sec:ind-gravity}

Now consider the quantization of our matrix model coupled to a 
scalar field.
In principle, the quantization is defined in terms of a (``path'') integral
over all matrices $Y^a$ and $\Phi$. 
In 4 dimensions, we can only perform perturbative computations
for the ``gauge sector'' encoded by $Y^a$, while the scalars can be integrated
out formally in terms of a determinant. Let us focus here on the
effective action obtained by integrating out the scalars,
\be
e^{-\Gamma_{\Phi}} = \int d\Phi e^{-S[\Phi]} 
\label{one-loop-action}
\ee
which for non-interacting scalar fields is given by 
\be
\Gamma_{\Phi} = \frac 12 \Tr \log \frac 12\Delta_{\tilde G} \, .
\label{trlog}
\ee
Here $\Delta_{\tilde G}$ is the Laplacian of a scalar field on
 the classical Riemannian manifold $(\cM,\tilde G^{ab}(y))$
with action \eq{scalar-action-geom}.
Later, we will consider an alternative interpretation as
Laplacian of a scalar field on $\R^4_{\bar \theta}$ coupled to an
adjoint $U(1)$ gauge field. 
In Feynman diagram language, \eq{trlog} will then
amount to the sum of all one-loop diagrams with arbitrary
numbers of external $A$-lines.
The subject of this paper is the comparison between these 2 different
computations of $\Gamma_{\Phi}$, once from the point of view of gravity
\eq{scalar-action-0}, and once from the point of view of NC $U(1)$ 
gauge theory \eq{scalar-action-A-tilde}.
This will provide an interpretation and understanding of the UV/IR
mixing for NC  gauge theory in terms of an induced gravitational
action (Einstein-Hilbert).

\paragraph{Induced gravity.}

We first focus on the geometric point of view. 
We want to compute the one-loop effective action $\Gamma_{\Phi}$
in terms of these classical geometrical data (which will later be expressed
in terms of classical $U(1)$ gauge fields $A_a(x)$). For this we write
\bea
\Tr \Big(\log\frac 12\Delta_{\tilde G}  - \log\frac 12\Delta_0\Big)
&\sim& -\Tr\int_{0}^{\infty} \frac{d\a}{\a}\,
(e^{-\a\frac 12\Delta_{\tilde G}} - e^{-\a\frac 12\Delta_0})\,\, \nn\\
&\equiv&\,\, -\Tr\int_{0}^{\infty} \frac{d\a}{\a}\,
\Big(e^{-\a\frac 12\Delta_{\tilde G}} - e^{-\a\frac 12\Delta_0 }\Big)\, e^{- \frac 1{\a\tilde \L^2}}
\label{TrLog-id}
\eea
where the small $\a$ divergence is regularized
using a UV cutoff $\tilde \L$, indicating that it is a cutoff for
$\Delta_{\tilde G}$.
Now we can use the heat kernel expansion, 
\be
\Tr e^{-\frac 12\a\Delta_{\tilde G}} \sim \sum_{n \geq 0}\, (\frac{\a}2)^{\frac{n-4}2}
\int_{\cM}\,  d^4 y\,\sqrt{|\tilde G_{ab}|}\,\,
 a_n(y,\Delta_{\tilde G})
\label{Seeley-deWitt}
\ee
where we can drop the measure $\sqrt{\det\tilde G}=1$. 
The $a_n(y,\Delta_{\tilde G})$ are known as Seeley-de Witt (or Duhamel)
coefficients, which for the action \eq{scalar-action-geom} are 
given by
\cite{Gilkey:1995mj} 
\bea
a_0(y) &=& \frac 1{16\pi^2}\, , \nn\\
a_2(y) &=& \frac 1{16\pi^2}\, \Big(\frac 16 R[\tilde G] \Big), \nn\\
a_4(y) &=& \frac 1{16\pi^2}\, \frac 1{360}\,  \(12 {{R_{;\mu}}}^\mu +
5 R^2 - 2 R_{\mu\nu} R^{\mu\nu} + 2
R_{\mu\nu\rho\sigma} R^{\mu\nu\rho\sigma}\), 
\label{SdW-coeff}
\eea
for the scalar case under consideration here (where $\cE=0$ and $\det
\tilde G=1$ in \cite{Gilkey:1995mj}).
Thus we obtain
\be
\Gamma_{\Phi} = \frac 1{16\pi^2}\, \int d^4 y \,\( -2\tilde \L^4 
- \frac 16 R[\tilde G]\, \tilde \L^2 + O(\log \tilde \L) \)\, .
\label{S-oneloop-scalar}
\ee
Recall that in general relativity, the term
$S_{const} \sim \int d^4 y\;\sqrt{\tilde G}\, \tilde \L^4$
corresponds to a cosmological constant, and 
its bad scaling behavior usually poses a major problem. Here we have
$\det \tilde G =1$, which suggests that this term
is essentially trivial. 
While this argument alone is not quite conclusive, 
we will find additional strong evidence that the 
cosmological constant problem is either absent or at least much milder
in the present framework. This would be great news, 
and will be discussed later. 
In particular, \eq{S-oneloop-scalar} suggests 
that the effective Newton constant is given
by the effective cutoff
\be
\frac 1{G} \sim \tilde \L^2 .
\label{Newton-cutoff}
\ee  
The curvature scalar $R[\tilde G]$ for the unimodular metric $\tilde G^{ab}$
can be expressed in terms of the curvature scalar $R[G]$ for $G^{ab}$
using
\bea
R[\tilde G] &=& \rho(y)\(R[G] +  3 \Delta_G \sigma-\frac 32\,G^{ab}
\partial_a\sigma \partial_b \sigma\)\, , \nn\\
\Delta_G \sigma &=&  G^{ab} \partial_a \partial_b \sigma - \Gamma^c \partial_c \sigma , \nn\\
\Gamma^a &=& G^{bc}\, \Gamma_{bc}^a\, , \nn\\
e^{-\sigma(y)} &=& \rho(y) = (\det G_{ab})^{1/4}\,.
\label{sigma-def}
\eea

\section{Geometry from $\mmu(1)$ gauge fields}
\label{sec:flat-expansion}

\subsection{Moyal-Weyl point of view.}

Let us now rewrite the geometric action 
\eq{scalar-action-0} in terms of the $\mmu(1)$ gauge fields on 
the flat Moyal-Weyl background $\R^4_\theta$ with generators $X^a$. 
This means that we consider ``small fluctuation''
\be
Y^a = X^{a} + \cA^a\, 
\label{cov-coord-1}
\ee
around the Moyal-Weyl generators $X^{a}$, which 
are solutions of the equations of motion \eq{eom} and
satisfy
\be
[X^a,X^b] = i \bar\theta^{ab}\, .
\label{Moyal-Weyl}
\ee
Here $\bar \theta^{ab}$
is a constant antisymmetric tensor. 
More precisely, we assume that the hermitian matrices
$\cA^a =\cA^a(X) \sim \cA^a(x)$ can be interpreted (at
least ``locally'') as smooth functions 
on $\R^4_{\bar \theta}$. 
Note that the effective geometry \eq{effective-metric}
 for the Moyal-Weyl plane is indeed
flat, given by
\bea
\bar g^{ab} &=& \bar\theta^{ac}\,\bar\theta^{bd} g_{cd}\, \nn\\
\tilde g^{ab} &=& \L_{NC}^4\, \bar g^{ab}   \nn\\
\bar\rho &=& |\det \bar g_{ab}|^{1/4} 
= (\det\bar\theta^{ab})^{-1/2} \equiv \L_{NC}^4 \, .
\label{effective-metric-bar}
\eea
Consider now the
change of variables
\be
\cA^a(x) = -\bar\theta^{ab} A_b(x)
\quad ( = i\bar\theta^{ab} i A_b)
\label{A-naive}
\ee
where $A_a$ is hermitian. 
Using
\be
[X^a + \cA^a,f] = i \bar\theta^{ab} (\frac{\partial}{\partial x^b} f
+ i [A_b,f]) \equiv  i \bar\theta^{ab} D_b f ,
\ee
the action \eq{scalar-action-0} can be written as 
\bea
S[\Phi] &=& \Tr\,\frac 12\bar\theta^{ab}\,\bar\theta^{a'c} g_{aa'}\,
 (\frac{\partial}{\partial x^b}\Phi + i[A_b,\Phi]) 
(\frac{\partial}{\partial x^c} \Phi + i[A_c,\Phi]) \nn\\
&=&  \int d^4 x\,\bar \rho\,\frac 12\bar g^{ab}\,
 (\frac{\partial}{\partial x^a}\Phi + i[A_a,\Phi]) 
(\frac{\partial}{\partial x^b} \Phi + i[A_b,\Phi]) \nn\\
&=&  \int d^4 x\, \frac 12\tilde g^{ab}\,  D_a\Phi D_b \Phi
= \int d^4 x\, \frac 12\, \Phi \Delta_A \Phi
\label{scalar-action-A-tilde}
\eea
where we define
\be
\Delta_A = - \tilde g^{ab}\, D_a D_b = -\bar\rho\, [Y^a,[Y_a,.]]
\label{Delta-A}
\ee
using \eq{effective-metric-bar}. 
Note that these formulas are exact if interpreted as 
noncommutative gauge theory on $\R^4_{\bar \theta}$, 
where $D_a = \frac{\partial}{\partial x^a} + i[A_a,.]$
is interpreted as covariant derivative 
with $\mmu(1)$ gauge field $A_a(x)$.

\subsection{Tensors and coordinate transformation 
 $y \to x$}
\label{sec:trafos}

Let us discuss the tensorial nature of the geometric 
objects and some associated subtleties.
The basic object is the dynamical Poisson structure
$\theta^{ab}(y)$ given by \eq{theta-def}, 
which is a rank 2 tensor in $y$ coordinates and
satisfies the Jacobi 
identity $\theta^{ad}(y)\partial_{y^d} \theta^{bc}(y) +
\mbox{cyclic} =0$.
Similarly, the effective metric \eq{G-U1} as well as $g_{ab}$ are 
tensors in $y$ coordinates. 
The coordinate system defined by the covariant coordinates $Y^a$
resp. $y^a$
is the natural one for the geometric point of view and hence for gravity.

On the other hand, using the $\mmu(1)$ gauge theory point of view
and the change of variables \eq{A-naive} we can express the 
Poisson tensor in terms of the  $\mmu(1)$ field strength  as
\be
i\theta^{ab}(Y) = [Y^a,Y^b] =  i\bar \theta^{ab} 
 - i\bar \theta^{ac}\bar \theta^{bd}\, \bar F_{cd} \, .
\label{theta-u1}
\ee
Here
$\bar F_{ab} = \partial_{x^a} A_b - \partial_{x^b} A_a + i[A_a,A_b]$
is a rank 2 tensor in $x$ coordinates on $\R^4_\theta$. 
This relates the Poisson tensor $\theta^{ab}(y)$ 
in $y$-coordinates with the 
field strength tensor $\bar F_{cd}$ in $x$-coordinates, where 
\be
y^a = x^a - \bar\theta^{ab} \, A_b \, .
\label{x-y-coords}
\ee
In order to avoid confusion 
we will denote all $x$-tensors with a bar in this section, and write
\be
\partial_a = \frac{\partial}{\partial y^a}, 
\qquad \bar\partial_a = \frac{\partial}{\partial x^a}\, ;
\ee
we will drop the bar in later sections if no confusion can arise.
Similarly, the induced metric $G^{ab}$ in $y$ coordinates
can be written in terms of the $\mmu(1)$ 
gauge fields as 
\be
G^{ab}(y) = \theta^{ac}(y) \theta^{b d}(y)\, g_{cd} 
 = (\bar \theta^{ac} 
  - \bar \theta^{ae}\bar \theta^{ch}\, \bar F_{eh}) 
(\bar \theta^{bd} - \bar \theta^{bf}\bar \theta^{dg}\,
\bar F_{fg})g_{cd} \, .
\label{G-U1}
\ee
Notice that while $\bar F_{ab}$ and $\bar \theta_{ab}$ are tensors in 
$x$ coordinates, $G^{ab}$ is a tensor in $y$ coordinates. 
Therefore if we want to compute e.g. Christoffel symbols, we must be 
careful to implement  the change of variables \eq{x-y-coords}, so that
\bea
\frac{\partial y^a}{\partial x^b} &=& \delta^a_b - \bar \theta^{ac}\, 
\bar\partial_{b} A_c = \delta^a_b - V^a_b, \nn\\
V_b^a(x) &=& \bar \theta^{ad}\,\bar\partial_b A_d(x)
\label{xy-trafo}
\eea
and
\be
\partial_a = \frac{\partial x^c}{\partial y^a}
\frac{\partial}{\partial x^c} 
\approx \bar\partial_a + V_a^c  \, \bar\partial_c 
\label{xy-trafo-deriv}
\ee
to leading order.
The Jacobian is given by 
\bea
\left|\frac{\partial y^a}{\partial x^b}\right| &=& 
\left|\d^a_b - V^a_b \, \right| 
= 1  - \bar\theta^{ac} \, \frac{\partial A_c}{\partial x^a} 
+ O(\bar\theta^2)  \nn\\
&=& 1 - \frac 12 \bar\theta^{ac} \, \bar F_{ac} + O(\bar\theta^2) \, .
\eea
This result holds\footnote{even if one would include the 2nd order term
in a Seiberg-Witten expansion \cite{Seiberg:1999vs};
however, using the SW expansion for $A$ is not appropriate here
because we want to compare with the results of the
non-expanded NC gauge theory.} even to $O(A^2)$ using \eq{det-expand}.

\paragraph{Metric and Poisson tensor}

Let us now consider the coordinate transformation $x^a \to y^a$
\eq{x-y-coords} for some of these tensors.
It is easy to see using \eq{theta-u1} that the Poisson tensor
$\theta^{ab}(y)$ on $y$-space is related to $\bar\theta^{ab}$
on $x$-space $\R^4_\theta$ using the diffeomorphism $x^a \to y^a$ 
to leading order in $\bar \theta$:
\be
\theta^{ab}(y) = (\one+V)^a_{a'}  (\one+V)^b_{b'} \bar\theta^{a'b'}\quad + O(\theta^3).
\ee
This means that $x^a$ can be interpreted as (local) 
Darboux coordinates for $\theta^{ab}(y)$, at least to the leading 
(semi-classical) order considered here. 
The relevance of Darboux coordinates for emergent gravity 
has been emphasized in \cite{Yang:2006hj} 
in the context of the DBI action; see also 
\cite{Cornalba:1999ah,Jurco:2000fb} for related discussion.
However, the effective metric $G^{ab}(y)$ \eq{effective-metric} is 
{\em not} obtained from either $g^{ab}$ or 
$\bar g^{ab}$ on $\R^4_\theta$ in this manner:
\be
G^{ab}(y) = \theta g \theta^T 
=(\one+V) \bar\theta (\one+V)^T g (\one+V) \bar\theta (\one+V)^T
\neq (\one+V) \bar g (\one+V)^T
\label{G-transform}
\ee
even to leading order. In particular,
$\det G^{ab} = (\det (\one+V))^4 \det \bar g \neq (\det (\one+V))^2 \det \bar g$.
This is essential, since otherwise $G^{ab}$ would be 
diffeo-equivalent to a constant metric and hence be flat\footnote{
In particular, this clarifies that the metrics $h_{ab}$ discussed
in the context of the DBI action \cite{Yang:2006hj}
do not contain the gravity described here, and 
are not equivalent to the effective metric 
\eq{effective-metric} which governs the matrix model.}.

\subsection{Rewriting the gravity action on $\R^4_{\theta}$}

We will now rewrite the  action $\Gamma_{\Phi}$ \eq{S-oneloop-scalar} 
in terms of gauge fields on $\R^4_{\theta}$ to $O(A^2)$.
The metric \eq{G-U1} is given by 
\bea
G^{ab} &=&
 (\bar \theta^{ac} - \bar \theta^{ae}\bar \theta^{ch}\, \bar F_{eh}) 
 (\bar \theta^{bd} - \bar \theta^{bf}\bar \theta^{dg}\,
\bar F_{fg})g_{cd} \, \nn\\
&=& \bar g^{ab} -h^{ab} 
\label{G-U2}
\eea
where
\bea
h^{ab} &=& - \bar g^{ad}\, \bar F_{df}\bar \theta^{fb}
- \bar \theta^{af}\, \bar F_{fd}\,\bar g^{bd}
 - \bar \theta^{ae}\, \bar F_{eh} \bar g^{hg} \,\bar F_{gf}\bar \theta^{fb}\nn\\
&=& - \bar g^{ad}\, \bar F_{df}\bar \theta^{fb}
- \bar \theta^{af}\, \bar F_{fd}\,\bar g^{bd} \, \, + O(A^2)\, .
\label{graviton-1}
\eea
This gives the linearized fluctuation resp. graviton in terms of the 
$\mmu(1)$ degrees of freedom. The linearized version 
was essentially found in \cite{Rivelles:2002ez}.

An immediate but important observation is that {\em the contributions 
linear in $A$ to 
the one-loop effective action \eq{S-oneloop-scalar} vanish identically}.
This holds because the metric fluctuations $h^{ab}$ are given by
derivatives of $A^a$, which vanish under the integral $\int d^4 y$ 
at $O(A)$ due to Stokes theorem.  
From the gauge theory point of view, this amounts to
the fact that
the vacuum $A=0$ is stable under quantization, i.e. the tadpole contributions vanish. 
This implies that 
{\em flat space $Y^a = X^a$ \eq{Moyal-Weyl} is a solution
of emergent gravity even after quantization}. 
Moreover, the same is expected to hold to all loops (using the gauge theory point of view). 
This observation is very significant, because it is strongly 
violated in the context of general relativity due to the induced 
cosmological constant term $\sim\int d^4 y\L^4 \sqrt{|g|}$\,: in GR,
flat space can only be preserved by very precise fine-tuning of the bare cosmological
constant. This is the infamous cosmological constant problem. 
We see here strong evidence that in
the context of emergent gravity from matrix models, this problem
appears to be
resolved or at least much milder. The basic reason is the  
constraint \eq{effective-metric} on the space of metrics.

In order to compute the determinant of the metric, the 
following form is more useful
\bea
G^{ab} &=& \bar g^{ar} \(\delta_r^b + \bar F_{rf} \bar \theta^{fb}
+ \bar g_{ra'} \bar \theta^{a'f}\, \bar F_{fd} \bar g^{db}\,
+ \bar g_{ra'}\bar \theta^{a'e}\, \bar F_{eh} \bar g^{hg} \bar F_{gf}\bar \theta^{fb} \)
\nn\\
&\equiv& \bar g^{ar} \(\delta_r^b + X_r^b\) .
\eea
To compute the determinant, we use
\be
\det(\one + X) 
= 1 + tr X + \frac 12 \((tr X)^2 - tr (X^2)\) \quad + O(X^3)  \,.
\label{det-expand}
\ee
From \eq{G-U2} we have
\bea
Tr X 
&=& \,- 2 \bar F_{rf} \bar \theta^{rf} 
- g^{e f} \bar F_{eh}\bar g^{hg} \bar F_{gf}
\eea
using 
$\bar\theta^{fb}\bar g_{ba}\bar\theta^{f'a} =  g^{ff'}$,
and one obtains
\bea
\frac{\det(G^{ab})}{\det(\bar g^{ab})} 
&=& 1 - 2 \bar F_{rf} \bar \theta^{rf} 
 + 2 (\bar F_{rf} \bar \theta^{rf})^2 
-  \bar F_{rf} \bar \theta^{fb}\bar F_{bf'} \bar \theta^{f'r} 
\eea
to $O(A^2)$.
This can be simplified further in the effective action 
where we can use partial integration at this order.
Using
\be
\int d^4 x\, \bar F_{rf} \bar \theta^{fb} \bar F_{bf'} \bar \theta^{f'r} 
= \int d^4 x\, \frac 12 \bar F_{r f'}  \bar \theta^{r f'}
 \bar F_{bf} \bar \theta^{bf}
\ee
we can write
\be
\det(G^{ab}) = \det(\bar g^{ab})\Big(1 -2 \bar F_{rf} \bar \theta^{rf} 
 + \frac 32 (\bar F_{rf} \bar \theta^{rf})^2 \Big) \, .
\ee
This gives
\be
e^\sigma = (\det G^{ab})^{1/4} 
= \det(\bar g^{ab})^{1/4}\,
\Big(1 - \frac 12 \bar F_{rf} \bar \theta^{rf} \, + O(\bar\theta^3)\Big)  ,
\label{e-sigma-id}
\ee 
hence
\be
\sigma =  \frac 14 \log\det(\bar g^{ab})\,
-\frac 12 \bar F_{rf} \bar \theta^{rf} 
- \frac 18 (\bar F_{rf} \bar \theta^{rf})^2 
\label{sigma-expand}
\ee
and
\be
\bar g^{ab}\partial_{a} \sigma \partial_{b} \sigma
= \frac 14 \bar g^{ab}\bar\partial_a (\bar F_{rf} \bar \theta^{rf})
  \bar\partial_b (\bar F_{r'f'} \bar \theta^{r'f'})\, .
\ee
noting that $\partial_{a} =\partial_{y^a} = \bar\partial_{a}$ to this order.
We also need
\be
\Delta_G \sigma = -G^{ab} \partial_a \partial_b \sigma + \Gamma^c \partial_c \sigma 
=  -\frac 1{\sqrt{G_{ab}}}\,
\partial_c\(\sqrt{G_{ab}}\, G^{cd} \partial_d \sigma\) .
\ee
To evaluate this, we need
\be
\sqrt{\det G_{ab}} = \sqrt{\det\bar g_{ab}} \,
(1 +\bar\theta^{ab} \bar F_{ab}) \quad + O(A^2) \, .
\ee
Therefore
\bea
\Delta_G \sigma &=& -\frac 1{\sqrt{G_{ab}}}\,
\partial_c\(\sqrt{G_{ab}}\, G^{cd} \partial_d \sigma\) \nn\\
&=& - (1 - \bar\theta^{ab} \bar F_{ab}) 
(\bar\partial_c + V_c^l \bar\partial_l)\(G^{cd} 
(1 + \bar\theta^{ab} \bar F_{ab})
(\bar\partial_d + V_d^l \bar\partial_l) \sigma \)  \quad + O(A^3)\nn\\
&=& -\bar\partial_c \(\bar g^{cd} V_d^l \bar\partial_l\sigma\) 
- V_c^l \bar\partial_l\(\bar g^{cd} \bar\partial_d \sigma\) 
- (1 - \bar\theta^{ab} \bar F_{ab}) \bar\partial_c \(G^{cd} 
(1 + \bar\theta^{ab} \bar F_{ab})\bar\partial_d \sigma\) \nn\\
&=& -\bar\theta^{ls}\bar\partial_l A_s  
\bar\partial^c \bar\partial_c \sigma
 - \bar\partial_c (G^{cd} \bar\partial_d \sigma) 
+ \bar\theta^{ab} \bar F_{ab} 
 \bar\partial^c \bar\partial_c \sigma \quad + \partial(O(A^2)) \nn\\
&=& -\frac 14 \bar\theta^{ab} \bar F_{ab} 
 \bar\partial^c \bar\partial_c  F_{de}\theta^{de}
+ \frac 12 \bar\partial^c \bar\partial_c F_{de}\theta^{de}
\eea
to the order required, omitting total derivatives 
of order $O(A^2)$ but not of order $O(A)$. 
Note that we used \eq{xy-trafo-deriv} in the 2nd line.
Therefore
\bea
3 \Delta_G \sigma 
- \frac 32 G^{ab}\partial_a \sigma \partial_b \sigma
&=& 
 \frac 32 \bar\partial^a \bar\partial_a\bar F_{rf} \bar\theta^{rf}
 - \frac 38 \bar F_{rf} \bar \theta^{rf}
   \bar\partial^a \bar\partial_a \bar F_{r'f'} \bar \theta^{r'f'}  .
\eea

\subsubsection{Ricci tensor and scalar curvature}

We recall the standard definitions:
\bea
\Gamma^e_{ab} & = & \frac12 G^{ce} \left( \partial_a G_{bc} + \partial_b G_{ca} - \partial_c G_{ab}
\right),\\
R^d_{abc} & = & \partial_b \Gamma^d_{ac} - \partial_c \Gamma^d_{ab} +
	\Gamma^s_{ac} \Gamma^d_{sb} - \Gamma^s_{ab} \Gamma^d_{sc},\\
R_{ac} & = & R^d_{adc},\\
R & = & G^{ab} R_{ab} \, .
\eea
These are tensors in $y$ coordinates.
The effective metric is 
\be
G^{ab} = \bar g^{ab} - h^{ab}
\ee
with $h^{ab}$ given by \eq{graviton-1}.
The inverse metric is given by
\be
G_{ab} = \bar g_{ab} + h_{ab} - h_{ac} h^c_{a} + \dots\,,
\ee
\be
\bar g_{ab} = (\bar g^{cd})^{-1}_{ab} = \bar \theta_{ac}^{-1} \bar\theta_{bd}^{-1}  g^{cd},
\ee
with $\bar \theta^{ac} \bar \theta_{cd}^{-1} = \delta^a_d$, yielding the identity
\be
\bar \theta^{ab} \bar g_{bc} \bar \theta^{dc} = g^{ad}
\,.
\ee
Indices are shifted with the metric $\bar g^{ab}$ and
its inverse. Therefore we obtain for the perturbation $h_{ab}$:
\be
\label{a2}
h_{ab} \equiv  \bar g_{ac} \bar g_{bd} h^{cd}
 =  - \bar g_{bu} \bar\theta^{uh} \bar F_{ha} - \bar g_{au} \bar\theta^{uh} \bar F_{hb} 
- \bar g_{ac} \bar g_{bd} \bar\theta^{df} \bar\theta^{ce} \bar g^{hg}
\bar F_{eh} \bar F_{fg} \, .
\ee
The terms of second order in $\theta$ in $h^{ab}$ 
and $h_{ab}$ contribute to the gravity action 
only to order higher than $\mathcal O(A^2)$ 
(using partial integration) and can be dropped here. 

When computing these quantities we must be careful to 
take into account the change of variables $x\to y$ \eq{xy-trafo-deriv}.
For the Christoffel symbols we get
\bea
\Gamma^i_{kl} & = & \frac12 \Big( 
\bar g^{im} (\bar\partial_l + V_l^c \, \bar\partial_c)h_{mk} 
+ \bar g^{im} (\bar\partial_k + V_k^c \, \bar\partial_c)  h_{ml} 
- \bar g^{im} (\bar\partial_m + V_m^c \, \bar\partial_c ) h_{kl} \nonumber\\
&& - h^{im} \bar\partial_l h_{mk} - h^{im} \bar\partial_k h_{ml} + h^{im} \bar\partial_m h_{kl} 
\quad + \partial(O(A^2))  \Big) \nn\\
&=:& \bar \Gamma^i_{kl} 
 +\bar g^{im} V_l^c \, \bar\partial_c h_{mk} 
+ \bar g^{im} V_k^c \, \bar\partial_c h_{ml} 
- \bar g^{im} V_m^c \, \bar\partial_c h_{kl}
\eea
to $O(A^2)$, defining the auxiliary object $\bar \Gamma^i_{kl}$.
To second order in $A$ 
the curvature tensor also picks up additional
terms upon rewriting $\partial_a$ with $\bar\partial_a$, and we have
\bea
R^d_{abc} & = & 
\partial_{b} \Gamma^d_{ac} - \partial_{c} \Gamma^d_{ab} +
	\Gamma^s_{ac} \Gamma^d_{sb} - \Gamma^s_{ab} \Gamma^d_{sc} \nn\\
 & = & \bar\partial_{b} \Gamma^d_{ac} + V_b^e\bar\partial_{e} \Gamma^d_{ac}
- \bar\partial_{c} \Gamma^d_{ab}  - V_c^e\bar\partial_{e} \Gamma^d_{ab}
+ \Gamma^s_{ac} \Gamma^d_{sb} - \Gamma^s_{ab} \Gamma^d_{sc} \nn\\
& = & \bar\partial_{b} \bar\Gamma^d_{ac} 
+ V_b^e\bar\partial_{e} \bar\Gamma^d_{ac}
 - \bar\partial_{c} \bar\Gamma^d_{ab} 
- V_c^e\bar\partial_{e} \bar\Gamma^d_{ab} 
 + \bar\Gamma^s_{ac} \bar\Gamma^d_{sb} - \bar\Gamma^s_{ab}\bar\Gamma^d_{sc}
\quad + \partial(O(A^2)) 
\eea
where we omit $O(A^2)$ terms which are total derivatives, since
we are only interested in the action to $O(A^2)$. Using partial integration
we can write
\bea
V_b^e\bar\partial_{e}\bar \Gamma^d_{ac}- V_c^e\bar\partial_{e}
\bar\Gamma^d_{ab}
&=& \bar \theta^{ef}\,\bar\partial_b A_f\bar\partial_{e}\bar
\Gamma^d_{ac}
- \bar \theta^{ef}\,\bar\partial_c A_f\bar\partial_{e}
\bar\Gamma^d_{ab}
=  \frac 12 \bar \theta^{ef}\bar F_{ef}\, (\bar\partial_b\bar\Gamma^d_{ac}
-  \bar\partial_c\bar\Gamma^d_{ab}) \nn\\
&=&  \frac 12 \bar \theta^{ed}\bar F_{ed}\, R^{(1)d}_{abc}
\eea
so that
\be
R^{d}_{abc} = \bar\partial_{b}\bar \Gamma^d_{ac} - \bar\partial_{c}\bar \Gamma^d_{ab}  
+ \bar\Gamma^s_{ac} \bar\Gamma^d_{sb} - \bar\Gamma^s_{ab} \bar\Gamma^d_{sc} 
 +  \frac 12 \bar \theta^{ef}\bar F_{ef}\, R^{(1)d}_{abc} 
\ee
to $O(A^2)$. 
We can compute the Riemann tensor as an expansion in the metric perturbation $h$
$$
R^{d}_{abc} = R^{(1)d}_{abc} + R^{(2)d}_{abc} + \dots \, .
$$
The first order term is
\be
R^{(1)d}_{abc}
 =  \frac12 \bar g^{dm} \bar\partial_b \bar\partial_a h_{mc} - \frac12 \bar g^{dm} \bar\partial_b \bar\partial_m h_{ac}
- \frac12 \bar g^{dm} \bar\partial_c \bar\partial_a h_{mb} + \frac12
\bar g^{dm} \bar\partial_c \bar\partial_m h_{ab} .
\ee
We can now compute the second order term of the Riemann tensor:
\bea
R^{(2)d}_{abc} & = & 
 \frac14 \bigg( \bar g^{sm} \bar\partial_c h_{ma} + \bar g^{sm} \bar\partial_a h_{mc} - \bar g^{sm} \bar\partial_m h_{ac}
\bigg)\,
\bigg( \bar g^{dn} \bar\partial_b h_{ns} + \bar g^{dn} \bar\partial_s h_{nb} - \bar g^{dn} \bar\partial_n h_{sb}\bigg)\nn\\
&& 
- \frac14 \bigg( \bar g^{sm} \bar\partial_b h_{ma} + \bar g^{sm} \bar\partial_a h_{mb} - \bar g^{sm} \bar\partial_m h_{ab}
\bigg)\,
\bigg( \bar g^{dn} \bar\partial_c h_{ns} + \bar g^{dn} \bar\partial_s h_{nc} - \bar g^{dn} \bar\partial_n h_{sc}
\bigg)\, \nn\\
&& +\frac 12 \bar \theta^{fe}\, \bar F_{fe} \, R^{(1)d}_{abc} 
\eea
omitting $O(A^2)$ terms which are total derivatives.
A suitable contraction gives the 2nd order Ricci tensor
\bea
R^{(2)}_{ac} & = &\frac14 \bar g^{sm} \bar g^{dn} \Big(
\bar\partial_c h_{ma} \bar\partial_d h_{ns} + \bar\partial_c h_{ma} \bar\partial_s h_{nd} - \bar\partial_c h_{ma} \bar\partial_n h_{sd} + \bar\partial_a h_{mc} \bar\partial_d h_{ns}  
\\
\nonumber
&&
+ \bar\partial_a h_{mc} \bar\partial_s h_{nd} - \bar\partial_a h_{mc} \bar\partial_n h_{sd} - \bar\partial_m h_{ac} \bar\partial_d h_{ns} - \bar\partial_m h_{ac} \bar\partial_s h_{nd}\\
\nonumber
&&
+ \bar\partial_m h_{ac} \bar\partial_n h_{sd} - \bar\partial_d h_{ma} \bar\partial_c h_{ns} - \bar\partial_d h_{ma} \bar\partial_s h_{nc} + \bar\partial_d h_{ma} \bar\partial_n h_{sc}  \\
\nonumber
&&
- \bar\partial_a h_{md} \bar\partial_c h_{ns} - \bar\partial_a h_{md} \bar\partial_s h_{nc} + \bar\partial_a h_{md} \bar\partial_n h_{sc} + \bar\partial_m h_{ad} \bar\partial_c h_{ns}\\
\nonumber
&&
+ \bar\partial_m h_{ad} \bar\partial_s h_{nc} - \bar\partial_m h_{ad} \bar\partial_n h_{sc}            
\Big) + \frac 12 \bar \theta^{fe}\, \bar F_{fe} \, R^{(1)}_{ac} .
\eea
Using partial integration this can be written as
\bea
R^{(1)}_{ac} & = & 
\bar\partial^s \bar\partial_{(a} h_{c)s} - \frac12 \bar\partial_a\bar\partial_c h - \frac12 \bar\partial^d\bar\partial_d h_{ac}\,,\\
R_{ac}^{(2)} & = & \frac14 h^{uv} \bar\partial_a \bar\partial_c h_{uv} - \frac12  h_{m(a}\bar\partial_{c)} \bar\partial^m h + \frac14 h_{ac} \bar\partial^c\bar\partial_c h
\\
\nonumber
&& + \frac12 h_{am} \bar\partial^m \bar\partial^d h_{dc} - \frac12 h^m_a \bar\partial^c\bar\partial_c h_{mc}
+ \frac 12 \bar \theta^{fe}\, \bar F_{fe} \, R^{(1)}_{ac}
\eea
where 
$$
\bar\partial_{(a} h_{c)s} = \frac12 (\bar\partial_{a} h_{cs} +
\bar\partial_{c} h_{as} ) \, .
$$
The Ricci scalar $R = G^{ab} R_{ab} = R^{(1)} + R^{(2)} 
=  \bar g^{ab}R_{ab}^{(1)} + \big(-h^{ab} R_{ab}^{(1)} +\bar g^{ab}
R_{ab}^{(2)}\big)$ 
contains the contributions
\bea
\label{a4}
R^{(1)} & = & \bar g^{ab} R^{(1)}_{ab} = 
\bar\partial^a \bar\partial^c h_{ac} - \bar\partial^c\bar\partial_c h,\\
\label{a3}
\bar g^{ab} R^{(2)}_{ab} 
&=& - \frac12 h_{ac} \bar\partial^a\bar\partial^c h + \frac14 h\bar\partial^c\bar\partial_c h  + \frac12 h^c_a \bar\partial^a \bar\partial^s h_{sc}
- \frac14 h^{ac} \bar\partial^c\bar\partial_c h_{ac}
+ \frac 12 \bar \theta^{fe}\, \bar F_{fe} \, R^{(1)} \, , \nn\\
h^{ab} R_{ab}^{(1)} &=& -\frac 12 \bar g^{ma} \bar g^{cr} \bar F_{rm} \bar\partial^2 \bar F_{ca}
-\frac 34 \bar F_{na}\bar \theta^{na}\bar\partial^c\bar\partial_c 
\bar F_{mf} \bar\theta^{mf} 
\eea
using \eq{hR-contraction} and partial integration, where 
\be
\bar\partial^2 = \bar\partial_a\bar\partial_b g^{ab}.
\label{partial2}
\ee
One easily computes in a similar way
\bea
h & = & \bar g^{cd} h_{cd}
 =  2 \bar \theta^{ab}\bar F_{ab}  + \mathcal O(A^2) \, , \nn\\
\bar\partial^a \bar\partial^b h_{ab}
&=& \bar\partial^a \bar\partial_a \bar\theta^{uh}\bar F_{uh} +
\mathcal O(A^2) 
\eea
so that
\be
R^{(1)} = - \bar\partial^a \bar\partial_a\bar \theta^{uh}\bar F_{uh}\, .
\label{a3-1}
\ee
The contributions to $\bar g^{ab} R^{(2)}_{ab}$  are of
the following form:
\bea
- \frac12 \int d^4x\, h_{ac} \bar\partial^a\bar\partial^c h & = & 
- \int d^4x\, \bar\theta^{ab} \bar F_{ab} \bar\partial^c\bar\partial_c
 \bar \theta^{cd} \bar F_{cd}\, ,\\
\frac14 \int d^4x\, h \bar\partial^c\bar\partial_c h 
& = & \int d^4x\, \bar\theta^{ab} \bar F_{ab}
\bar\partial^c\bar\partial_c \bar\theta^{cd} \bar F_{cd} \, ,\\
 \int d^4x \, \frac12 h^c_a \bar\partial^a \bar\partial^s h_{sc} 
  -\frac14  h^{ab} \bar\partial^c\bar\partial_c h_{ab}
&=& \int d^4x \left(
\frac 18 \bar \theta^{ab} \bar F_{ab} \bar\partial^c\bar\partial_c \bar \theta^{cd} \bar F_{cd}
+ \frac14 \bar g^{ah} \bar g^{mr}\bar F_{hm}  \bar\partial^2 \bar F_{ra}
\right) \, . \nn\\
\label{id-5}
\eea
Collecting these terms, we find
\bea
\nonumber
\int d^4 x R^{(2)} 
& = & \int d^4 x \left(
\frac{3}8 \bar \theta^{ab} \bar F_{ab} \bar\partial^c\bar\partial_c \bar \theta^{cd} \bar F_{cd}
+ \frac14 \bar g^{ma}  \bar g^{hr} \bar F_{hm} \bar\partial^2 \bar F_{ra}
\right) \, .
\eea
We can finally write down the one-loop induced action
\eq{S-oneloop-scalar}
\bea
\Gamma_{\Phi} &=& \frac 1{16\pi^2}\, \int d^4 y \, \(-2\tilde \L^4 - 
\frac 16 \rho(y)\(R[G] +  3 \Delta_G \sigma-\frac 32\,G^{ab}
\partial_a\sigma \partial_b \sigma\)\tilde \L^2 + O(\log \tilde \L) \)\nn\\
 &=& \frac 1{16\pi^2}\, \int d^4 y \, \Bigg(-2\tilde \L^4 - \frac 16 \rho(y)\Big(
 \frac 12\bar\partial^a \bar\partial_a\bar \theta^{uh}\bar F_{uh} 
+ \frac14 \bar g^{ma}  \bar g^{hr} \bar F_{hm} \bar\partial^2 \bar F_{ra} 
\Big)  \tilde \L^2 \,\,
+ O(\log \tilde \L) \Bigg) \, .\nn\\
\eea
In order to compare this with the gauge theory computation 
on $\R^4_\theta$, we have to rewrite this action on $x$ space.
There is a  subtlety concerning the cutoffs: 
$\tilde \L$ is the effective cutoff for $\Delta_{\tilde G}$,
which acts on the Hilbert space of function with 
inner product $(f,g) = \int d^4 y\, f(y)^* g(y)$.
On the gauge theory side, we have an effective  
cutoff $\L$ for $\Delta_A = \bar\rho\, [Y^a,[Y_a,.]]$ \eq{Delta-A}
which acts on the Hilbert space of function with 
inner product $\left< f,g\right> = Tr f^\star g = 
\int d^4 y\, \rho(y)f(y)^* g(y)$.
To understand the relation between $\Delta_{\tilde G}$ 
and $\Delta_A$ we can write the action in 
2 equivalent ways (adding a mass term for clarity)
\bea
S[\Phi] &=& \int d^4 y\, \frac 12\, \Big( -\Phi 
\partial_{y^a}\big(\tilde G^{ab}(y)  \partial_{y^b}\Phi\big) 
+\rho(y) m^2 \Phi^2 \Big)
= \int d^4 y\,\frac 12\, \Big(\Phi \Delta_{\tilde G} \Phi 
+\rho(y) m^2 \Phi^2 \Big) \nn\\
&=&  Tr \frac 12\,\Big(-\Phi [Y^a,[Y_a,\Phi]] + m^2 \Phi^2\Big) = 
\int d^4 y\,\frac 12\,\frac{\rho(y)}{\bar\rho} 
\Big( \Phi \Delta_A \Phi+\bar\rho m^2 \Phi^2\Big) \, .
\label{action-2}
\eea
This means that 
\be
\Delta_{\tilde G} = \frac{\rho(y)}{\bar\rho}\, \Delta_A
\ee
in $y$ coordinates (the gauged kinetic term $\Delta_A$ is usually written in
$x$ coordinates, but expressed in $y$ coordinates here), 
which reflects the use of the rescaled 
metric $\tilde G$ \eq{metric-unimod}. Since we implement the cutoffs
using a Schwinger parameter as in \eq{TrLog-id} resp.
\eq{schwinger},
this means that the effective cutoffs are related as 
\be
\tilde \L^2 = \frac{\rho(y)}{\bar\rho}\, \L^2 \, .
\ee
Such a ``local cutoff'' makes sense provided
$\frac{\rho(y)}{\bar\rho}\,$ varies only on large scales resp. small momenta $p \ll \L$,
which is indeed our working assumption. The same conclusion is found using a 
Pauli-Villars regularization (or in a softly broken supersymmetric setting),
where the mass $m \sim \L$ in \eq{action-2} plays the role of the cutoff.
Noting that
\bea
d^4 y \rho(y) &=& d^4 x \bar\rho \, ,\nn\\
\frac{\rho(y)}{\bar\rho}\, &=& \Big(\frac{\det G_{ab}}{\det\bar g_{ab}}\Big)^{1/4}
= 1 + \frac 12 \bar\theta^{uh}\bar F_{uh} + \frac 14 (F_{ab}\bar\theta^{ab})^2
\eea
we finally obtain
\bea
\Gamma_{\Phi} 
 &=& \frac 1{16\pi^2}\, \int d^4 x \,
\Bigg(-2\L^4(1+ \frac 14 (F_{ab}\bar\theta^{ab})^2 ) -
 \bar \rho\,\frac{\L^2}{24} \Big(
 \bar\theta^{ab} \bar F_{ab}\bar\partial^a \bar\partial_a\bar \theta^{uh}\bar F_{uh} 
+ \bar g^{ma}  \bar g^{hr} \bar F_{hm} \bar\partial^2 \bar
F_{ra}\Big) \nn\\
&& \qquad\qquad + O(\log \L)\,\, + O(A^3) \Bigg) 
\label{effective-geom-A2}
\eea
dropping terms which vanish under the integral.
This has precisely the form obtained from UV/IR mixing 
\eq{effective-action-UVIR} in the gauge
theory approach.

\section{Comparison with UV/IR mixing}
\label{sec:UV-IR-mixing}

In this section, we compare the result of the previous section 
with the one-loop  effective action from
the gauge theory point of view. The result is of course the same, 
but the gauge-theory computation sheds new light on the conditions
to which extent the semi-classical analysis of the previous section is valid,
and allows to compute corrections to \eq{S-oneloop-scalar}.
In particular, we find indeed 
- as predicted in \cite{Steinacker:2007dq} - 
that the well-known but thus far mysterious UV/IR mixing terms in 
the effective action for NC gauge theory are precisely given by the 
induced gravity (=Einstein-Hilbert plus $\L^4$ term)
action \eq{S-oneloop-scalar}, in a suitable IR regime.
More precisely, this holds provided
\be
p\, \L  < \L_{NC}^2
\label{semiclassical-regime}
\ee
which amounts to ``mild'' UV/IR mixing.
In particular, we need an explicit, physical momentum cutoff $\L$,
which should typically be of order $\L \leq \L_{NC}$ for the above
regime to be physically interesting. In that case, 
\eq{semiclassical-regime} follows from
\be
p < \L_{NC},
\label{cutoff}
\ee
which is very reasonable range of validity for the classical gravity action.
Such a cutoff $\L \leq \L_{NC}$ could be provided e.g. by 
a softly or spontaneously broken supersymmetric completion of the model,
which will be discussed later. Dimensional regularization, on the 
other hand, does not appear to be useful here.

Even though the one-loop effective action for NC gauge theory
has been computed in many places 
\cite{Hayakawa:1999zf,Minwalla:1999px,Matusis:2000jf,VanRaamsdonk:2001jd,Khoze:2000sy},
the results given in the literature
are not sufficiently precise in the IR limit for our purpose. 
The point is that we need to analyze carefully the IR regime of
the well-known effective cutoff $\L_{eff}(p)$ \eq{lambda-eff} 
for non-planar graphs as $p \to 0$, keeping $\L$ fixed. 
In other words, we consider the regime where the non-planar 
diagrams 
almost coincide with the planar diagrams, and keep the leading 
NC corrections. This corresponds precisely to the leading semiclassical 
terms inherent in e.g. \eq{derivation}.
This regime has not been considered in
previous attempts to explain UV/IR mixing,  e.g. in terms of
exchange of closed string modes \cite{Armoni:2001uw}.

To understand the need for an explicit cutoff $\L$,
 it is instructive to consider a
regularization of $\R^4_{\theta}$ given e.g. by fuzzy tori
 \cite{Ambjorn:1999ts}
or fuzzy $\C P^2$ resp. $S^2\times S^2$ \cite{Grosse:2004wm}. One then typically  
finds in addition to the NC scale $\L_{NC}$ explicit UV- and IR cutoffs,
\be
\L_{IR} = \sqrt{\frac{1}{N\theta}} \quad \ll \quad
 \L_{NC} = \sqrt{\frac{1}{\theta}} \quad \ll \quad 
\L_{UV} = \sqrt{\frac{N}{\theta}}, 
\label{fuzzy-scales}
\ee
where $N$ is related to the dimension $\cN$ of the matrices (typically $\cN
\sim N^2$ in 4D). Note that $\L_{UV} = \infty$ for $\R^4_{\theta}$.
This type of fuzzy regularization does {\em not} suffice here;
if we set $\L = \L_{UV}$, then \eq{semiclassical-regime}
together with the condition \eq{fuzzy-scales}
for the semiclassical regime would give $p \leq \L_{IR}$,
which leaves no room for interesting physics.  
Then the geometrical action \eq{S-oneloop-scalar}
would have to be replaced by a strongly non-commutative 
one, which would presumably lead to new phase transitions and
new phenomena such as striped phases \cite{Gubser:2000cd,Bietenholz:2006cz}.
In this paper, we focus on the semi-classical 
regime \eq{semiclassical-regime}.

\subsection{One-loop computation}

\begin{figure}[h]
\begin{center}
\includegraphics[scale=0.7]{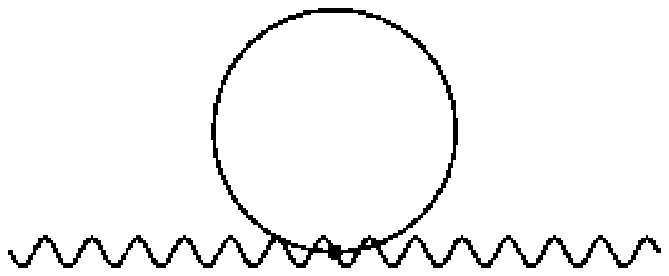}
\end{center}
\caption{}
\label{fig:2}
\end{figure}

Consider now the action  \eq{scalar-action-A-tilde}
for a scalar coupled to the $\mmu(1)$  gauge field, written 
in Moyal-Weyl space so that $\bar \theta^{ab} = const$. 
We can cast the action in the form  
\bea
S[\Phi] &=& \int d^4 x\, \frac 12
\tilde g^{ab} (\bar\partial_a + ig[A_a,.])\Phi 
(\bar\partial_b + ig[A_b,.])\Phi + \frac 12 m^2 \Phi^2
\,\,=\,\, S_0[\Phi] + S_{int}[\Phi], \nn\\
S_0[\Phi] &=& \int d^4 x\, \frac 12  \tilde g^{ab}
\bar\partial_a \Phi \bar\partial_b \Phi + \frac 12 m^2 \Phi^2
\eea
with the unimodular metric $\tilde g^{ab}$ defined in 
\eq{effective-metric-bar}. We introduced an explict coupling constant
$g$ and a mass $m$ here\footnote{following the conventions 
of section \ref{sec:flat-expansion} we should actually write 
$\bar\rho m^2$, but we absorb $\bar\rho$ in $m^2$ here.}.
The propagators will involve the metric $\tilde
g^{ab}$, and we will write
\newpage
\bea
k\cdot k &\equiv& k_i k_j\, \tilde g^{ij} \, ,  \nn\\
k^2 &\equiv& k_i\, k_j g^{ij}  
\label{norm-notation}
\eea
from now on.
We need the $O(A^2)$ contribution to the 
1-loop effective action obtained by integrating out the scalar
$\Phi$:
\be
e^{-\Gamma_\Phi} = \left< \exp\Big(-\int d^4 x\, (ig\bar\partial_a \phi
[A_b,\phi]\tilde g^{ab} 
- \frac{g^2}2 [A_a,\phi][A_b,\phi]\tilde g^{ab})\Big)\right> \, .
\ee
While this has been considered several times in the 
literature, the known results are 
not accurate enough for our purpose, i.e. 
in the regime $p^2, \L^2<\L_{NC}^2$ where the semiclassical geometry is 
expected to make sense. 
We therefore compute carefully
\bea
\Gamma_{\Phi} &=& \frac 12 \Tr \log \Delta_0 
-\frac{g^2}2 \left<\int d^4 x\,i\bar\partial_a \phi[A_b,\phi]
\tilde g^{ab} \int d^4 y\,i\bar\partial_a \phi[A_b,\phi]\tilde g^{ab}\right> \nn\\
&& - \frac{g^2}2 \left<\int d^4 x\, [A_a,\phi][A_b,\phi]\tilde g^{ab}\right> \nn\\
&=& \frac 12 \Tr \log \Delta_0 +\Gamma_{\Phi}^{(1)} + \Gamma_{\Phi}^{(2)} .
\eea
\begin{figure}[h]
\begin{center}
\includegraphics[scale=0.7]{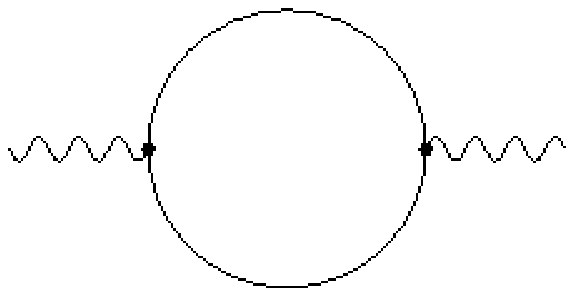}
\end{center}
\caption{}
\label{fig:1}
\end{figure}
The contribution from diagram a) in figure \ref{fig:1} is given by
\bea
\Gamma_{\Phi}^{(1)}  &=& -\frac{g^2}2 \int \frac{d^4 p}{(2\pi)^4}\, 
A_{a'}(p) A_{b'}(-p) \tilde g^{a'a} \tilde g^{b'b}
\int \frac{d^4 k}{(2\pi)^4}\,
\frac{4k_a k_b + 2 k_a p_b + 2 p_a k_b + p_a p_b}
{(k\cdot k +m^2)((k+p)\cdot(k+p) +m^2)}\, \nn\\
&& \(1-e^{i k_i \theta^{ij} p_j}\) \nn\\
&=& \Gamma_{\Phi}^{(1),P} + \Gamma_{\Phi}^{(1),NP}\, .
\label{Gamma-1}
\eea
The contribution from diagram b) in figure \ref{fig:2} is given by
\bea
\Gamma_{\Phi}^{(2)}  &=& g^2\int \frac{d^4 p}{(2\pi)^4}\, 
A_{a'}(p) A_{b'}(-p) \tilde g^{ab}
\int \frac{d^4 k}{(2\pi)^4}\,
\frac{1}{(k\cdot k+m^2)}\, \(1-e^{i k_i \theta^{ij} p_j}\) \nn\\
&=& \Gamma_{\Phi}^{(2),P} + \Gamma_{\Phi}^{(2),NP}\, .
\label{Gamma-2}
\eea
As a small check,
note that the planar and nonplanar parts cancel for $\theta=0$,
because $\Phi$ is in the adjoint and decouples for $\theta=0$.

As explained above, in order to compare this with the induced
gravity action \eq{S-oneloop-scalar}
we must regularize these divergent integrals
using a momentum cutoff $\L$. We will do this here 
``by hand'' using a suitable cutoff for the Schwinger parameters.
This procedure is applicable for all loops.
We will show moreover in Appendix B that 
precisely this prescription is obtained by carefully implementing
the same regularization as in the geometrical action \eq{TrLog-id}.
Therefore we should expect to find precise agreement with 
\eq{S-oneloop-scalar}, which is indeed the case.

\subsection{Some integrals}

We use the Schwinger representation for propagators 
\bea
 \frac{1}{k\cdot k + m^2} &=& \int_0^\infty d\a\, e^{-\a(k\cdot k+m^2)}, \nn\\
\frac{1}{(k\cdot k + m^2)^2} &=& \int_0^\infty d\a\,\a e^{-\a(k\cdot k+m^2)},\nn\\
\eea
where  $k\cdot k \equiv \tilde g^{ij} k_i k_j$,
and put a small mass as an IR regulator. 
The UV cutoff is conveniently implemented using 
the following regularization 
\be
\frac{1}{k\cdot k} \to \int_0^\infty d\a\, e^{-\a k\cdot k - \frac{1}{\L^2 \a}} 
\label{schwinger}
\ee
which removes the UV singularity at $\a=0$.
For this regularization we need the following integrals:
\bea
\int_0^\infty d\a\, \frac 1\a\, e^{-\a m^2 - \frac{1}{\L^2 \a}}
 &=& 2 K_0\Big(2 \sqrt{\frac{m^2}{\L^2}}\Big)  
\,=\,  - 2 \(\gamma + \log(\sqrt{\frac{m^2}{\L^2}})\) 
\, + O\Big(\frac{m^2}{\L^2}\log(\frac{\L}{m})\Big) \, ,\nn\\
\int_0^\infty d\a\, \frac 1{\a^2}\, e^{-\a m^2 - \frac{1}{\L^2 \a}}
 &=& 2 \sqrt{\L^2 m^2}\,K_1(2 \sqrt{\frac{m^2}{\L^2}}) \nn\\
&=&    \L^2 - 2 m^2\log(\sqrt{\frac{\L^2}{m^2}}) 
 + m^2(2\gamma -1)  + O\Big(\frac{m^4}{\L^2}\log(\frac{\L}{m})\Big)\, , \nn\\
\int_0^\infty d\a\, \frac 1{\a^3}\, e^{-\a m^2 - \frac{1}{\L^2 \a}}
 &=& 2 \L^2 m^2\,K_2(2 \sqrt{\frac{m^2}{\L^2}})  \nn\\
&=&   \L^4 - m^2\L^2 + m^4 \log(\sqrt{\frac{\L^2}{m^2}}) 
 - m^4(\gamma - \frac 34) + O\Big(\frac{m^6}{\L^2}\log(\frac{\L}{m})\Big) \nn\\
\label{K-asymptotics}
\eea
where $\gamma$ is the Euler constant. We
will drop finite terms which vanish for $m \to 0$, apart
from those needed to have dimensionless arguments.

\subsection{$\Gamma_{\Phi}^{(1)}$}

\begin{figure}[h]
\begin{center}
\includegraphics[scale=0.4]{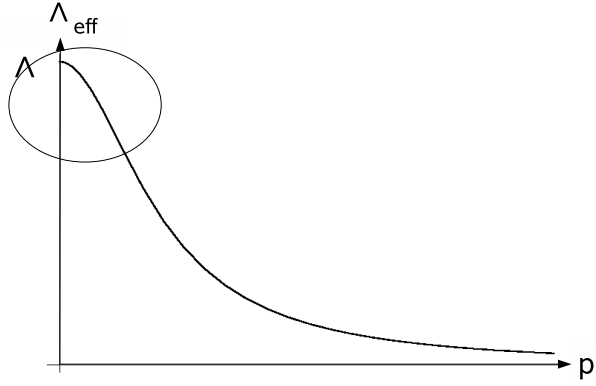}
\end{center}
\caption{relevant IR regime of $\L_{eff}(p)$}
\label{fig:L-eff}
\end{figure}

It is convenient to write \eq{Gamma-1} 
using a Feynman parameter
\bea
\frac{1}{(k\cdot k+m^2)((k+p)\cdot(k+p) +m^2)} &=& \int_0^1 dz\, 
\frac 1{(l\cdot l + z(1-z) p\cdot p + m^2)^2} \nn\\
&=& \int_0^1 dz\, 
 \int_0^\infty d\a\, \a e^{-\a(l\cdot l + z(1-z) p\cdot p + m^2)}
\eea
where
\be
l = k + z p \,.
\ee
We need
\bea
&& \int \frac{d^4 k}{(2\pi)^4}\; \frac{P(k)}{(k\cdot k +m^2)((k+p)\cdot(k+p) +m^2)}
    (1-e^{i k_i \theta^{ij} p_j}) \nn\\
&=& \int \frac{d^4 k}{(2\pi)^4}\; P(k) \int_0^1 dz\, 
 \int_0^\infty d\a\, \a e^{-\a(l\cdot l + z(1-z) p\cdot p + m^2)-\frac 1{\L^2\a}}
 (1-e^{i k \theta p}) \nn\\
&=&  \int_0^1 dz\,  \int_0^\infty d\a\,\a\, e^{-\a (z(1-z) p\cdot p + m^2)-\frac 1{\L^2\a}}
\int \frac{d^4 l}{(2\pi)^4}\; P(l-zp) 
   (e^{-\a l\cdot l} -e^{-\a (l_i l_j  + i l_i \frac{\tilde p_j}{\a}) \tilde g^{ij}}) \nn\\
&=&  \int_0^1 dz\,  \int_0^\infty d\a\,\a\, e^{-\a (z(1-z) p\cdot p +
  m^2)-\frac 1{\L^2\a}}  \nn\\
&& \qquad  \int \frac{d^4 l}{(2\pi)^4}\; \(P(l-zp) e^{-\a l\cdot l}
-  P(l-zp+i \frac{\tilde p}{2\a}) 
        e^{-\a l\cdot l- \frac{\tilde p\cdot\tilde p}{4\a}}\) \, .\nn
\eea
We completed the square and shifted the integration 
$l \to l + i \frac{\tilde p}{2\a}$
in the last expression, where
\be
\tilde p_i = \tilde g_{ij} \tilde p^j = \tilde g_{ij} \theta^{jk} p_k \,.
\ee
For our purpose, $P(k)$ is a polynomial which is at most quadratic.  
We have
\bea
\int d^4 l\; e^{-\a l\cdot l} &=& 
  \pi^2\frac 1{\a^2}, \qquad
\int d^4 l\; l_i e^{-\a l\cdot l} = 0, \nn\\
\int d^4 l\; l_i l_j e^{-\a l\cdot l} 
&=& \frac 12 \pi^2\tilde g_{ij}\, \frac{1}{\a^3}
\eea
since $\det \tilde g^{ij}=1$. Therefore 
\bea
&& \int \frac{d^4 k}{(2\pi)^4}\, 
\frac{4k_a k_b + 2 k_a p_b + 2 p_a k_b + p_a p_b}
{(k\cdot k+m^2)((k+p)\cdot(k+p)+m^2)}\, (1-e^{i k \theta p}) \nn\\
&=& \frac{1}{16\pi^2}\, \int_0^1 dz\, \int_0^\infty d\a\,e^{-\a
  (z(1-z) p\cdot p + m^2)-\frac 1{\L^2\a}} 
\, \Big(p_a p_b \frac 1{\a} (1-4z+4z^2)+ 2 g_{ab} \frac 1{\a^2} \nn\\
&& -\Big( p_a p_b(1-4z+4z^2) \frac 1{\a}
 + i (p_a \tilde p_b + \tilde p_a p_b) \frac{1 -2z}{\a^2}
 - \tilde p_a\tilde p_b\frac 1{\a^3}
+ 2\tilde g_{ab} \frac 1{\a^2} \Big) e^{- \frac{\tilde p\cdot\tilde p}{4\a}} \Big) \nn\\
&=& \frac{1}{16\pi^2}\, \int_0^1 dz\, \Big(2 p_a p_b(1-4z+4z^2) \,
\Big(K_0\big({\textstyle 2\sqrt{\frac{z(1-z)p\cdot p + m^2}{\L^2}}}\big)
 - K_0\big({\textstyle 2\sqrt{\frac{z(1-z)p\cdot p + m^2}{\L_{eff}^2}}}\big) \Big)\nn\\
&& - \big(i (p_a \tilde p_b + \tilde p_a p_b) (1 -2z)
+2 \tilde g_{ab}\big)2\sqrt{(z(1-z) p\cdot p+ m^2)\L_{eff}^2}\,
K_1\big({\textstyle 2\sqrt{\frac{z(1-z)p\cdot p + m^2}{\L_{eff}^2}}}\big) \nn\\
&& + 2\tilde p_a\tilde p_b 
 (z(1-z)p\cdot p + m^2)\L_{eff}^2\,
 K_2\big({\textstyle 2\sqrt{\frac{z(1-z)p\cdot p + m^2}{\L_{eff}^2}}}\big) \nn\\
&& + 2 \tilde g_{ab}\sqrt{(z(1-z)p\cdot p + m^2)\L^2} \,
2 K_1\big({\textstyle 2\sqrt{\frac{z(1-z)p\cdot p + m^2}{\L^2}}}\big)\Big)\Big). \nn\\
\eea
This is exact. Note that the term proportional to $(p_a \tilde p_b + \tilde p_a p_b)$
vanishes identically under $\int dz$ (as it must, because it
is not gauge invariant).
Here
\bea
\L_{eff}^2 &=& { 1 \over 1/\Lambda^2 + \frac 14\tilde p \cdot \tilde p} 
= { 1 \over 1/\Lambda^2 + \frac 14 \frac{p^2}{\L_{NC}^4}} 
= \L_{eff}^2(p) 
\label{lambda-eff}
\eea
is the ``effective'' cutoff for non-planar graphs,
noting that
\be
\tilde p\cdot \tilde p = \tilde p_i\tilde p_j\tilde g^{ij}
= \bar\theta^{ii'} \bar\theta^{jj'} p_{i'} p_{j'}\tilde g_{ij}
= g^{i'j'}\,p_{i'} p_{j'} \, (\det \bar g^{ab})^{1/4}
\equiv \frac{p^2}{\L_{NC}^4}\, .
\ee
Now we consider the IR regime  
\be
\frac{p^2 \Lambda^2}{\L_{NC}^4} < 1 \, ,
\label{IR-regime}
\ee
see figure \ref{fig:L-eff}.
Then both $\L$ and $\L_{eff}$ are large, and we can 
use the asymptotic expansions \eq{K-asymptotics}.
Thus
\bea
&& \int \frac{d^4 k}{(2\pi)^4}\, 
\frac{4k_a k_b + 2 k_a p_b + 2 p_a k_b + p_a p_b}
{(k\cdot k+m^2)((k+p)\cdot(p+k)+m^2)}\,(1-e^{i k \theta p})  \nn\\
&& \qquad = \frac{1}{16\pi^2}\int_0^1 dz\, 
 \Big(- p_a p_b(1-2z)^2\,\log\Big(\frac{\L_{eff}^2}{\L^2}\Big)\nn\\
&& \qquad\qquad - 2\tilde g_{ab}\Big(\L_{eff}^2 - \L^2
+ 2(z(1-z)p\cdot p + m^2)\log(\sqrt{\frac{\L^2}{\L_{eff}^2}})+ ...\Big)\nn\\
&&\qquad\qquad + \tilde p_a\tilde p_b 
 \Big(\L_{eff}^4 - (z(1-z)p\cdot p + m^2)\L_{eff}^2 \nn\\
 &&\qquad\qquad - (z(1-z)p\cdot p + m^2)^2 \log \Big(\sqrt{{\textstyle\frac{z(1-z)p\cdot p 
  +  m^2}{\L_{eff}^2}}}\Big) + ...\Big)\Big) \nn\\
&& \qquad = \frac{1}{16\pi^2} \Big(-\frac 13 (p_a p_b - \tilde g_{ab}\,p\cdot p)
 \,\log\Big(\frac{\L_{eff}^2}{\L^2}\Big)
 - 2g_{ab}\Big(\L_{eff}^2 - \L^2\Big)\nn\\
&&\qquad\qquad + \tilde p_a\tilde p_b 
 \Big(\L_{eff}^4 - \frac 16 p\cdot p\L_{eff}^2 
 +\frac{(p\cdot p)^2}{1800}\, 
(47-30\log({\textstyle\frac{p\cdot p}{\L_{eff}^2}}))
 \Big)\Big) \, ,\nn
\eea
dropping finite terms vanish for $m \to 0$.

\subsection{$\Gamma_{\Phi}^{(2)}$}

Here we have
\bea
\int \frac{d^4 k}{(2\pi)^4}
\frac{(e^{i k_i \theta^{ij} p_j} -1)}{(k\cdot k+m^2)}\, 
&=& \int \frac{d^4 k}{(2\pi)^4}\;  
 \int_0^\infty d\a\, e^{-\a(k\cdot k + m^2)-\frac 1{\L^2\a}}
 (e^{i k \theta p} -1) \nn\\
&=& \int_0^\infty d\a\, e^{-\a m^2-\frac 1{\L^2\a}}
\int \frac{d^4 k}{(2\pi)^4}\; 
(e^{-\a (k - i \frac{\tilde p}{2\a})\cdot(k - i \frac{\tilde p}{2\a})
- \frac{\tilde p\cdot\tilde p}{4\a}} - e^{-\a k\cdot k})\nn\\
 &=& \frac{1}{16\pi^2}\, 2\Big(\sqrt{m^2\L_{eff}^2}K_1(2\sqrt{\frac{m^2}{\L_{eff}^2}}) 
 - \sqrt{m^2\L^2}K_1(2\sqrt{\frac{m^2}{\L^2}})\Big)\nn\\
&\sim& \frac{1}{16\pi^2}\,\Big(\L_{eff}^2 - \L^2 
+ 2 m^2\log(\sqrt{\frac{m^2}{\L_{eff}^2}})  
  - 2 m^2\log(\sqrt{\frac{m^2}{\L^2}}  \Big) \nn\\
&=& \frac{1}{16\pi^2}\,\(\L_{eff}^2 - \L^2 \)
\eea
up to terms of order 
$O(m^2 \frac{p^2 \L^2}{\L_{NC}^4})$ which vanish for $m \to 0$.

\subsection{Effective action}

Combining the above results and using
\bea
\bar F_{ab}(p) \bar F_{a'b'}(-p) \tilde g^{aa'} \tilde g^{bb'} 
&=& 2\(p_a p_{a'} A_b(p) A_{b'}(-p) - p_a p_{b'} A_b(p) A_{a'}(-p)\)  
\tilde g^{aa'} \tilde g^{bb'} \nn\\
(\theta\bar F(p)) (\theta\bar F(-p)) 
&=&  4 (\tilde p^a A_a(p))(\tilde p^b A_b(-p)) 
\eea
we obtain the induced action 
\bea
\Gamma_{\Phi} &=& \Gamma_{\Phi}^{(1)} + \Gamma_{\Phi}^{(2)} \nn\\
&=& - \frac{g^2}2\frac{1}{16 \pi^2}\,\int \frac{d^4 p}{(2\pi)^4}\, \, 
 \Big(-\frac 1{6}  \bar F_{ab}(p) \bar F_{a'b'}(-p) 
\tilde g^{a'a} \tilde g^{b'b}
 \,\log(\frac{\L^2}{\L_{eff}^2}) \nn\\
&&\qquad + \frac 14 (\theta\bar F(p)) (\theta\bar F(-p))
 \Big(\L_{eff}^4 - \frac 16 p\cdot p\, \L_{eff}^2 
 +\frac{(p\cdot p)^2}{1800}\, 
(47-30\log({\textstyle\frac{p\cdot p}{\L_{eff}^2}})) 
 \Big)\Big)  \nn\\
\label{ind-action-gauge}
\eea
up to finite terms vanish for $m \to 0$.
This is manifestly gauge invariant.

To make contact with the geometrical action \eq{effective-geom-A2},
we use the expansions
\bea
\L_{eff}^2 &=& \Lambda^2 - p^2 \frac{\Lambda^4 }{4\L_{NC}^4} +... \, ,
\nn\\
\L_{eff}^4 &=& \Lambda^4 - p^2 \frac{\Lambda^6 }{2\L_{NC}^4} +... \, ,
\nn\\
\log(\frac{\L^2}{\L_{eff}^2})&=& 
\frac 14 \frac{p^2 \L^2}{\L_{NC}^4} + ... 
\label{lambda-eff-expand}
\eea
which are valid in the IR regime \eq{IR-regime}.
Now assume first

\underline{$\L \ll \L_{NC}$}.

In this case 
\be
\L^6\frac{p^2}{\L_{NC}^4} = \frac{\L^4}{\L_{NC}^4} \L^2 p^2 
\,\,\ll\,\, \L^2  p^2 \,\,\sim\,\,  \L^2  p\cdot p 
\label{correction-cond}
\ee
so that we can replace
\be
\L_{eff}^4 - \frac 16 p\cdot p \L_{eff}^2 
\,\,\sim \,\,\L^4 - \frac 16p\cdot p \L^2\, .
\ee
Then the leading terms are 
\bea
\Gamma_{\Phi} &\sim&  -\frac{g^2}2\frac{1}{16 \pi^2}\,
\int \frac{d^4 p}{(2\pi)^4}\, 
 \Big(\frac{\L^4}4  (\theta\bar F(p)) (\theta\bar F(-p)) 
-\frac{\L^2}{24} \bar F_{ab}(p)\, \bar F_{a'b'}(-p) 
\frac{p^2}{\L_{NC}^4}
\tilde g^{a'a} \tilde g^{b'b} \nn\\
&& \qquad\qquad\quad - \frac{\L^2}{24}(\theta\bar F(p))(\theta\bar F(-p)) 
p\cdot p \Big)  \nn\\
&=& -\frac{g^2}2\frac{1}{16 \pi^2}\, \int d^4 x\, 
 \Big(\frac{\L^4}4 (\theta\bar F) (\theta\bar F)
+ \bar \rho\,\frac{\L^2}{24} 
 \Big( \bar F_{ab}\,\bar\partial^2 \, \bar F_{a'b'} \bar g^{a'a} \bar g^{b'b}
 + (\theta\bar F)\bar\partial^a\bar\partial_a(\theta\bar F) \Big) \nn\\
&& \qquad\qquad + O(\log(\L))\,\, + \rm{finite} \Big)\, ,  \nn
\label{effective-action-UVIR}
\eea
using  \eq{effective-metric-bar} 
and the notation \eq{partial2} and \eq{norm-notation}.
This coincides precisely with the effective action obtained
from the induced gravity action \eq{effective-geom-A2}, 
upon replacing $\L^2 \to 2\L^2$ and absorbing $g$ in $A$. 
We will show moreover in Appendix B that 
precisely this prescription is obtained by carefully implementing
the same regularization as in the geometrical action \eq{TrLog-id}.

If desired, the $O(\log(\L))$ and finite terms can be obtained from 
\eq{ind-action-gauge}. They involve higher derivative expressions,
and should reproduce \eq{SdW-coeff}; we will not verify this here.
Finally, the constant term 
$-\frac 1{8\pi^2}\int d^4 x\, \L^4$ in \eq{effective-geom-A2} 
represents the phase space volume of states with 
$\Delta_0 < \L$, but is physically irrelevant here. 
It can be obtained if desiresd using the same regularization as above 
(inserting a test function $f$ for mathematical rigor)
\bea 
 -\frac 12 \Tr \log (\frac 12\Delta_0) f
&=& \frac 12 Tr \int_0^\infty d\a\, \frac 1{\a} 
e^{-\a \frac 12\Delta_0}\, e^{ - \frac{1}{\L^2 \a}} \, f \nn\\
&=&-\frac 12 \int \frac{d^4 k}{(2\pi)^4}\, f(k)
\int_0^\infty d\a\,\frac 1{\a} 
e^{-\a \frac 12 k\cdot k  - \frac{1}{\L^2 \a}}\nn\\
&=&-\frac 12 \L^4\, \int \frac{d^4 k}{(2\pi)^4}\, f(k\L)
\int_0^\infty d\a\,\frac 1{\a} 
e^{-\a \frac 12 k\cdot k  - \frac{1}{\a}}\nn\\
&=&-\frac 1{8\pi^2} \L^4\, \int d^4 k\, f(k)\, \d^{(4)}(k)
=-\frac 1{8\pi^2} \L^4\, \int d^4 x\,f(x) 
\eea
using a standard rescaling argument,
since $\int d^4 k\int_0^\infty d\a\,\frac 1{\a}  e^{-\a \frac 12 k\cdot k  -
  \frac{1}{\a}} = 4\pi^2$.
This can also be seen from a theorem by H. Weyl
\cite{weyl} on the asymptotics of eigenvalue distributions.

Let us summarize the main result from the gauge theory 
point of view. We obtained a
simple geometrical explanation of the ``strange'' IR behavior of NC
gauge theory. In particular, the hitherto mysterious
$\L^4$ divergence reflects the 
leading term in the density of states 
of a Laplacian coupled to a background metric, 
and the $\L^2$ terms correspond precisely to the Einstein-Hilbert action. 
From the gravity point of view, perhaps the most remarkable point is 
that the $\L^4$ term is compatible here
with the existence of flat space, unlike in general relativity.

Even though we focused on the IR regime 
here in order to make contact with the classical geometry, 
there is nothing which prevents us from considering the effective
action also for higher energies. 
In particular, we obtain a first correction 
beyond the classical Einstein-Hilbert
by assuming

\underline{$\L \sim \L_{NC}$}.

If $\L \ll \L_{NC}$ does not hold,
we cannot neglect $\L^6\frac{p^2}{\L_{NC}^4}$ compared to
$\L^2  p\cdot p$ \eq{correction-cond}, leading to
correction terms in the gravitational action compared 
with the classical result.
For $\L \sim \L_{NC}$, we find 
\bea
\Gamma_{\Phi} &\sim&
-\frac{g^2}2\frac{1}{16 \pi^2}\, \int d^4 x\, 
 \Big(\frac{\L^4}4  (\bar F \theta)^2
+ \frac{\L^6}4 (\bar F \theta) 
\frac{\bar\partial^2}{\L_{NC}^4} (\bar F \theta) \nn\\
&&\qquad \qquad \quad + \bar \rho\,\frac{\L^2}{24} 
 \Big( \bar F_{ab}\,\bar\partial^2 \, \bar F_{a'b'}  
\bar g^{a'a}  \bar g^{b'b}
 +  (\bar F \theta) \bar\partial^a\bar\partial_a 
(\bar F \theta)\Big) \Big) \, . \nn
\eea
The additional term can be written e.g. as
\be
\frac{1}4 (\bar F \theta) 
\bar\partial^2 (\bar F \theta)
 \sim \sigma \bar\partial^2 \sigma
\ee
using \eq{sigma-expand}. Note that its $\L^6$ coefficient is somewhat
misleading, since it is simply a term in the Taylor expansion 
of $\L_{eff}^4 \leq \L^4$. 
Finally, terms quartic in momentum will arise in the action as soon as
we violate the IR regime \eq{IR-regime}, due to higher-order terms in
the Taylor expansion of $\L_{eff}$. In that case, one 
would enter an entirely new regime which is suspected to lead to 
new types of phenomena and phase transitions
(``striped phase'') such as those 
discussed in \cite{Gubser:2000cd,Bietenholz:2006cz}. 
Whether this can be understood in suitably
generalized geometrical terms remains to be seen.

\section{Conclusion and outlook}

In this paper we perform a nontrivial check for the basic result in
\cite{Steinacker:2007dq}, which gives an interpretation of the $\mmu(1)$ sector 
of NC gauge theory (in the 
Matrix-Model formulation) in terms of gravity. 
This provides an explanation of the
well-known ``strange'' IR behavior of the NC gauge theory
at the quantum level in terms
of an induced Einstein-Hilbert action. 
We verified this prediction explicitly
by comparing the one-loop effective action induced by a scalar field
from the geometrical and from the gauge theory point of view.
We are able to match and explain the precise form 
of the UV/IR mixing terms of NC gauge theory to $O(A^2)$
in the IR limit, including the leading $\L^4$ divergence. 
This confirms the geometric form \eq{S-oneloop-scalar} 
for the one-loop effective action
in the semiclassical limit, as 
well as the formula \eq{effective-metric} for the
effective metric. In particular, \eq{S-oneloop-scalar} provides much more
information than the 2-particle sector usually considered
in NC gauge theory.

The geometric interpretation 
in terms of an Einstein-Hilbert action
applies in the semi-classical IR regime.
We therefore consider the IR limit $p \to 0$ of the gauge theory
where the well-known ``effective cutoff'' for the non-planar diagrams
$\L_{eff}^2 = \frac 1{\frac 1{\L^2} 
+ \frac 14 \tilde p\cdot\tilde p}$ can be expanded in a Taylor
series around $p=0$. The geometrical picture 
resp. the Einstein-Hilbert action turns out to be 
valid for momenta $p \L < \L_{NC}^2$.
This allows a physically reasonable range of momenta
$p \leq \L_{NC}$, provided we
assume a cutoff $\L < \L_{NC}$. This is a scaling
regime which apparently has not been considered
in the literature up to now.
Moreover, we obtain correction terms to the Einstein-Hilbert 
action for an extended range of momenta resp. cutoff.

While the gravitational point of view
 provides an interpretation and understanding
of UV/IR mixing, it does not by itself render the theory 
renormalizable. If we remove the cutoff $\L$ resp. 
set $\L = \L_{UV}$, the induced gravitational action 
diverges and cannot be absorbed by adjusting the bare parameters.
However, this insight does suggest a way how to make these
models well-defined and physically meaningful, by ensuring that 
there really is a cutoff $\L \leq \L_{NC}$. One 
natural way to achieve this is to make the model 
supersymmetric, with spontaneously or softly broken 
supersymmetry. This should result in a well-defined
NC quantum field theory, which contains (emergent)
quantized gravity as an intrinsic part. 
There is indeed an obvious candidate for such a model, namely
the IKKT model \cite{Ishibashi:1996xs}, interpreted as $N=4$ NCSYM
in 4 dimensions.
This model can be modified e.g. by adding soft SUSY breaking terms. 
Interestingly enough this provides a 
direct link with string theory, 
which may provide further new insights; see 
also \cite{Aoki:1999vr,Ishibashi:2000hh} for related work.

Emergent NC gravity therefore provides a new, 
rather direct link between gravity and gauge
theory in 4 dimensions. Some aspects of this relation 
have been discussed in \cite{Yang:2006hj}.
We make this relation explicit by 
expressing the curvature scalar 
$R[\tilde G]$ in terms of $U(1)$ gauge fields. This could be
extended to arbitrary 
$n$-graviton scattering amplitudes, as long as the
class of geometries covered by \eq{effective-metric} is appropriate.  
Whether this can be 
related to other proposed relations between gravity and 
gauge theory such as \cite{Bern:2002kj} remains to be seen. 
UV/IR mixing for NC gauge theory can now be understood as a
relation between the gravitational IR regime of the model
and the UV regime which is interpreted in terms of 
gauge theory.
A possible relation with different attempts \cite{Azuma:2002af}
to identify gravity within the IKKT matrix model 
is unclear at present. 

While some considerations in this paper are mathematically justified
only in the Euclidean case, the main ideas apply equally well to the
case of Minkowski signature. Therefore the issue of Wick rotation
should be re-investigated using the specific assumptions 
in our context,
in particular the restriction to the IR regime \eq{IR-regime}.
Furthermore, we only consider the $U(1)$ case in this paper
for simplicity
(which means pure gravity); the extension to nonabelian gauge 
fields was given in \cite{Steinacker:2007dq}. 
Nevertheless the one-loop quantization of a nonabelian
scalar field should be worked out explicitly in the same regime, 
since it will not only lead to an induced gravity action but also 
to the (standard) renormalization of the $SU(n)$ gauge fields. 
We recall that the UV/IR mixing is restricted to the $U(1)$ sector.

Finally, the present computation is a first step
towards a complete one-loop computation of the effective action,
which then allows to study the physical aspects 
and viability of emergent 
gravity. In particular, the cosmological constant problem
is expected not to be present or much milder,
since flat space remains to be a vacuum solution
at one loop. This should 
provide enough motivation for further work.

\subsection*{Acknowledgments}

H.S. would like to thank M. Buric, J. Madore, P. Schupp, 
and H-S. Yang for discussions, and in particular 
to L. Freidel and  L. Smolin for  discussions 
and hospitality at the Perimeter Institute for theoretical physics. 
The work of H.S. was supported by the FWF Project P18657, and 
the work of M.W. was supported in part by the FWF Project P20017 
and in part by the Erwin-Schr\"odinger Institute for mathematical physics.

\section{Appendix A: Computation of $h^{ab} R_{ab}$}

We compute the leading contribution to $h^{ab} R_{ab}$.
To first order in $A$, one finds
\be
R^{(1)}_{ab} =  
- \frac12 \bar\theta^{uh} \bar\partial_a \bar \partial_b \bar F_{uh} 
+ \frac12 \bar\theta_a^{\,\,\,h} \bar\partial^c \bar \partial_h \bar F_{cb}
+ \frac12 \bar\theta_b^{\,\,\,h} \bar\partial^c \bar \partial_h \bar F_{ca}\,.
\ee
Contraction with the fluctuation of the metric gives
\bea
&&  \!\!\!\!\!\!\!\!  2 \int d^4x h^{ab} R^{(1)}_{ab} \nn\\
 & = &  
-\int d^4 x
( \bar g^{am} \bar F_{mn} \bar \theta^{nb} + \bar g^{bm} \bar F_{mn} \bar \theta^{na} )
( \bar\theta_a^{\,\,\,h} \bar\partial^c \bar \partial_h \bar F_{cb} 
+ \bar\theta_b^{\,\,\,h} \bar\partial^c \bar \partial_h \bar F_{ca}
- \bar\theta^{uh} \bar\partial_a \bar \partial_b \bar F_{uh}  )
\nn\\
& = & -2\int d^4 x
\Big(
 -  \bar g^{ma} g^{nh} \bar F_{mn} \bar\partial^c  \bar\partial_h \bar F_{ca}
+  \bar \theta^{na} \bar\theta^{mf} \bar F_{mn} \bar \partial^c \bar \partial_f \bar F_{ca}
+ \bar \theta^{na} \bar\theta^{mf} \bar F_{cn} \bar \partial^c \bar \partial_a \bar F_{fm}
\Big) \nn\\
&=&  -\int d^4 x
\Big(\bar g^{ma} \bar g^{cr} \bar F_{rm} \bar\partial^2 \bar F_{ca}
+\frac 32  \bar F_{na}\bar \theta^{na}\bar\partial^c\bar\partial_c \bar
F_{mf} \bar\theta^{mf} \Big) 
\label{hR-contraction}
\eea
where we have used (up to 2nd order in $A$)
\begin{itemize}
\item
\bea
  \!\!\!\!\!\!\!\! \int d^4 x\, \bar g^{ma} g^{nh} \bar F_{mn} \bar\partial^c
 \bar\partial_h \bar F_{ca} 
& = &  -\int d^4 x\, \bar g^{ma} g^{nh} \bar g^{cr} \bar\partial_r\bar F_{mn} 
 \bar\partial_h \bar F_{ca} \nn\\
 &=& \int d^4 x\, \bar g^{ma} g^{nh} \bar g^{cr} 
(\bar\partial_m\bar F_{nr} +  \bar\partial_n\bar F_{rm} ) 
 \bar\partial_h \bar F_{ca} \nn\\
 &=& \int d^4 x\,  g^{nh} \bar g^{am} 
\bar F_{nm} \bar\partial^c \bar\partial_h \bar F_{ca}
 - \bar g^{ma}  \bar g^{cr}  \bar F_{rm} \bar\partial^2 \bar F_{ca} \nn\\
\eea
hence
\be
2 \int d^4 x\, \bar g^{ma} g^{nh} \bar F_{mn} \bar\partial^c
 \bar\partial_h \bar F_{ca} 
 =  - \int d^4 x\,\bar g^{ma} \bar g^{cr} \bar F_{rm} \bar\partial^2 \bar F_{ca}
\ee

\item 
\bea
2 \int d^4 x \bar \theta^{na} \bar\theta^{mf} \bar F_{mn} \bar \partial^c \bar \partial_f \bar F_{ca}
& = & - 2 \int d^4 x\bar \theta^{na} \bar\theta^{mf} \bar \partial_f \bar F_{mn} \bar \partial^c \bar F_{ca}
\nn\\
&& \hspace{-2cm}
= \int d^4 x\bar \theta^{na} \bar\theta^{mf} \bar \partial_n \bar F_{fm} \bar \partial^c \bar F_{ca}
= \int d^4 x\bar \theta^{na} \bar\theta^{mf} \bar \partial^c \bar F_{fm} \bar \partial_n \bar F_{ca}
\nn\\
&& \hspace{-2cm}
= \frac12 \int d^4 x\bar \theta^{na} \bar\theta^{mf} \bar \partial^c \bar F_{fm} \bar \partial_c \bar F_{na}
= \frac12 \int d^4 x\bar \theta^{na} \bar\theta^{mf} 
\bar\partial^c\bar\partial_c \bar F_{mf} \bar F_{na} \nn
\eea

\item 
\bea
2 \int d^4 x \bar \theta^{na} \bar\theta^{mf} \bar F_{cn} \bar \partial^c \bar \partial_a \bar F_{fm}
& = & - 2 \int d^4 x \bar \theta^{na} \bar\theta^{mf} \bar \partial_a \bar F_{cn} \bar \partial^c \bar F_{fm}
\nn\\
&& \hspace{-2cm}
= - \int d^4 x \bar \theta^{na} \bar\theta^{mf} \bar \partial_c \bar F_{an} \bar \partial^c \bar F_{fm}
=   \int d^4 x \bar \theta^{na} \bar\theta^{mf} 
\bar\partial^c\bar\partial_c \bar F_{na} \bar F_{mf} \nn
\eea

\end{itemize}

\section{Appendix B: Regularization for the gauge theory}

Consider the one-loop effective action in terms of 
the gauge field $A$:
$$
\Gamma_{\Phi} = 
\frac 12 \Tr \Big(\log\frac 12\Delta_{A}  - \log\frac 12\Delta_0\Big)
\equiv\,\, -\frac 12\Tr\int_{0}^{\infty} \frac{d\a}{\a}\,
\Big(e^{-\a\frac 12\Delta_{A}} - e^{-\a\frac 12\Delta_0 }\Big)\, 
e^{- \frac 1{\a\L^2}} 
$$
where the small $\a$ divergence is regularized as in \eq{TrLog-id}
using a UV cutoff $\L$.
To obtain the expansion in $A$ we use the Duhamel formula 
(cf. \cite{Grosse:2007dm})
\bea
e^{-\a H}  - e^{-\a H^0} &=&  -  \int_0^\a d t_1 
e^{-t_1 H^0} V e^{-(\a-t_1) H^0} 
\nonumber \\
&&+  
\int_0^\a d t_1 \int_0^{t_1} d t_2 
e^{-t_2 H^0} V e^{-(t_1-t_2) H^0} V e^{-(\a-t_1) H^0} + \dots
\eea
where $H = H_0 + V$. 
Now
\bea
&& \int_0^\infty \frac{d\a}{\a}\,\int_0^\a d t_1 \int_0^{t_1} d t_2 
Tr\Big(e^{-t_2 H^0} V e^{-(t_1-t_2) H^0} V e^{-(\a-t_1) H^0}\Big)\, 
e^{- \frac 1{\a\L^2}} \nn\\
&& = \int_0^\infty \frac{d\a}{\a}\, \int_0^{\a} d t'\,\int_{t'}^\a dt_1 
Tr\Big( V e^{-t' H^0} V e^{-(\a-t') H^0}\Big)\, 
e^{- \frac 1{\a\L^2}} \nn\\
&& =  \int_0^\infty \frac{d\a}{\a}\, \int_0^{\a} d t'\,(\a-t')
Tr\Big( V e^{-t' H^0} V e^{-(\a-t') H^0}\Big)\, 
e^{- \frac 1{\a\L^2}} \nn\\
&& = \int_0^\infty \frac{d\a}{\a}\, \int_0^{\a} d t''\,t''
Tr\Big(V e^{-t'' H^0} V e^{-(\a-t'') H^0} \Big)\, 
e^{- \frac 1{\a\L^2}} 
\eea
where $t' = t_1-t_2$ and $t'' = \a-t'$. Combining the two last 
lines we obtain
\bea
&& \int_0^\infty \frac{d\a}{\a}\,\int_0^\a d t_1 \int_0^{t_1} d t_2 
Tr\Big(e^{-t_2 H^0} V e^{-(t_1-t_2) H^0} V e^{-(\a-t_1) H^0}\Big)\, 
e^{- \frac 1{\a\L^2}} \nn\\
&& = \frac {1}2\,\int_0^\infty d\a\, \int_0^{\a} d t''
Tr\Big(V e^{-t'' H^0} V e^{-(\a-t'') H^0} \Big)\, 
e^{- \frac 1{\a\L^2}} 
\eea
so that
\bea
\Gamma_{\Phi} & = &
\frac{1}2\, \int_0^\infty d\a\, \Tr (V e^{-\a H^0}) e^{- \frac 1{\a\L^2}}
- \frac{1}4 \int_0^\infty d\a \int_0^\a dt'\,\,
    \Tr \Big(V e^{-t'H^0} V e^{-(\a-t')H^0}\Big) e^{- \frac 1{\a\L^2}}
    \, .
\nonumber
\eea
Here $\Tr$ denotes the trace of operators acting 
on the scalar field on $\R^4_{\theta}$, which is conveniently written
in momentum basis  $\Phi(x) = \int \frac{d^4p}{(2\pi)^4}\, \Phi(p) \,
e^{i p_a x^a}$. 
Using 
\be
(H_0)_{p,q} = \frac 12 (\Delta_0)_{p,q} 
= \frac 12 p^2\, \delta_{p,q}
\ee
and 
\be
e^{i k x} e^{i l x} = e^{-\frac i2 k \theta l}\, e^{i (k+l) x} 
\ee
 on $\R^4_{\theta}$, the interaction term becomes
\bea
\langle\phi|V|\phi\rangle &=& \int d^4 x\, \Big(ig\bar\partial_a \phi
[A_b,\phi]\tilde g^{ab} 
- \frac{g^2}2 [A_a,\phi][A_b,\phi]\tilde g^{ab}\Big) \nn\\
&=& \int \frac{d^4 p\,d^4 q }{(2\pi)^4(2\pi)^4}\, 
 \Big( \phi(p)^*
 \Big(\frac 12 g\, (p_a+q_a) A_b(p-q)\tilde g^{ab} 2 i \sin(-\frac 12 p
 \theta q)\Big) \phi(q) \nn\\
&&\!\!\! + \int \frac{d^4 l}{(2\pi)^4}\, 
 \phi(p)^* \Big(\frac{g^2}2\,A_a(l) A_b(p-l-q)\tilde g^{ab} 2\sin(\frac 12 l \theta p)
2\sin(-\frac 12 (p-l) \theta q) \Big) \phi(q) \Big) \nn\\
&\equiv& \int \frac{d^4 p\,d^4 q}{(2\pi)^4(2\pi)^4}\, 
\Phi(p)^* V_{p,q} \Phi(q) \, .
\eea
Keeping only quadratic expressions in $A$, we obtain
\bea
\Tr (V e^{-\a H^0}) = \frac{g^2}2\,
\int \frac{d^4 l}{(2\pi)^4} A_a(l) A_b(-l)\tilde g^{ab} 
\int \frac{d^4 p}{(2\pi)^4}\,
(4\sin^2(\frac 12 l \theta p)) e^{-\frac 12 p^2\,\a} 
\eea
and
\bea
\Tr \Big(V e^{-t'H^0} V e^{-(\a-t')H^0}\Big)
&=& \int \frac{d^4 p\,d^4 q }{(2\pi)^4 (2\pi)^4} 
 \Big(g\, (p_a+q_a) A_b(p-q)\tilde g^{ab}  i \sin(-\frac 12 p
 \theta q)\Big) \, e^{-\frac 12 q^2\, t'} \nn\\
&& \Big( g\, (p_a+q_a) A_b(q-p)\tilde g^{ab}  i \sin(-\frac 12 q
 \theta p)\Big) \, e^{-\frac 12 p^2\, (\a-t')} \nn\\
&=& g^2\int \frac{d^4 p}{(2\pi)^4} A_b(p)\tilde g^{ab}
A_{b'}(-p)\tilde g^{a'b'}
\int\frac{d^4 q }{(2\pi)^4} (p_a+2q_a)(p_{a'}+2q_{a'}) \nn\\
&& \qquad \sin^2(-\frac 12 p\theta q)
\, e^{-\frac 12 q^2\, t'-\frac 12 (p+q)^2\, (\a-t')} \, .
\eea
Altogether we obtain
\bea
\Gamma_{\Phi} & = & 
g^2 \int \frac{d^4 p}{(2\pi)^4} A_a(p) A_b(-p)\tilde g^{ab} 
\int \frac{d^4 k}{(2\pi)^4}\,
\sin^2(\frac 12 p \theta k) \int_0^\infty d\a\, 
e^{-\frac 12 k^2\,\a - \frac 1{\a\L^2}}  \nn\\
&&
- \frac{g^2}4 \int \frac{d^4 p}{(2\pi)^4} A_b(p)\tilde g^{ab}
A_{b'}(-p)\tilde g^{a'b'}
\int\frac{d^4 k }{(2\pi)^4}\, (p_a+2k_a)(p_{a'}+2k_{a'}) 
 \sin^2(\frac 12 p\theta k)\nn\\
&& \qquad \int_0^\infty d\a \int_0^\a dt'\, 
\, e^{-\frac 12 k^2\, t'-\frac 12 (p+k)^2\, (\a-t')} 
e^{- \frac 1{\a\L^2}} \, .
\eea
The first term contains the expected Schwinger parameter,
and the second term involves
$$
 \int_0^\infty d\a \int_0^\a dt'\, 
\, e^{-\frac 12 k^2\, t'-\frac 12 (p+k)^2\, (\a-t')
- \frac 1{\a\L^2}} 
 =  \int_0^\infty d\a \int_0^\a d\b\, 
\, e^{-\frac 12 \a (k + \frac{\b}{\a}\, p)^2 
 + \frac 12 (\frac{\b^2}{\a} -\b ) p \cdot p\,
- \frac 1{\a\L^2}} 
$$
where $\b = \a - t'$. Define
\be
l\equiv k + z \, p , \qquad z = \frac{\b}{\a}\,
\ee
so that
$$
\int_0^\infty d\a \int_0^\a dt'\, 
\, e^{-\frac 12 k^2\, t'-\frac 12 (p+k)^2\, (\a-t')} 
e^{- \frac 1{\a\L^2}} 
 = 4 \int_0^\infty d\a' \, \a' \int_0^1 d z\, 
\, e^{- \a'(l^2 + z(1-z)\, p\cdot p)- \frac 1{2\a' \L^2}}
$$
where 
\be
\a'=\a/2 .
\ee
Now
\be
 \sin^2(\frac 12 k\theta p) = \frac 12 (1-\cos(k\theta p))
\sim \frac 12(1-e^{i k\theta p})
\ee
provides the distinction between planar and non-planar diagrams;
the latter replacement is justified under the integrals in the present
context.
Thus we end up exactly with the regularization in section \ref{sec:UV-IR-mixing},
\bea
\Gamma_{\Phi} & = & 
 g^2 \int \frac{d^4 k}{(2\pi)^4} A_a(p) A_b(-p)\tilde g^{ab} 
\int \frac{d^4 k}{(2\pi)^4}\,(1-e^{i k\theta p}) \int_0^\infty d\a'\, 
e^{- k^2\,\a' - \frac 1{2\a'\L^2}}  \nn\\
&&
- \frac{g^2}2 \int \frac{d^4 p}{(2\pi)^4} A_b(p)\tilde g^{ab}
A_{b'}(-p)\tilde g^{a'b'}
\int\frac{d^4 k }{(2\pi)^4}\, (p_a+2k_a)(p_{a'}+2k_{a'}) 
 (1-e^{i k\theta p})\nn\\
&& \qquad \int_0^\infty d\a' \a' \int_0^1 d z\, 
\, e^{- \a'(l^2 + z(1-z)\, p\cdot p)- \frac 1{2\a' \L^2}} \,,
\eea
with $\L^2$ replaced by $2\L^2$.


\begin{thebibliography}{99}




\bibitem{Steinacker:2007dq}
  H.~Steinacker,
  ``Emergent Gravity from Noncommutative Gauge Theory,''
  JHEP {\bf 12}, (2007) 049; 
 [arXiv:0708.2426 [hep-th]].

\bibitem{Rivelles:2002ez}
  V.~O.~Rivelles,
  ``Noncommutative field theories and gravity,''
  Phys.\ Lett.\  B {\bf 558} (2003) 191
  [arXiv:hep-th/0212262].


\bibitem{Yang:2006hj}
  H.~S.~Yang,
  ``Instantons and emergent geometry,''
  arXiv:hep-th/0608013; 
  H.~S.~Yang,
  ``Emergent gravity from noncommutative spacetime,''
  arXiv:hep-th/0611174;
  H.~S.~Yang,
  ``On The Correspondence Between Noncommuative Field Theory And Gravity,''
  Mod.\ Phys.\ Lett.\  A {\bf 22} (2007) 1119
  [arXiv:hep-th/0612231].

\bibitem{Muthukumar:2004wj}
  B.~Muthukumar,
  ``U(1) gauge invariant noncommutative Schroedinger theory and gravity,''
  Phys.\ Rev.\  D {\bf 71} (2005) 105007
  [arXiv:hep-th/0412069].


\bibitem{Matusis:2000jf}
  A.~Matusis, L.~Susskind and N.~Toumbas,
  ``The IR/UV connection in the non-commutative gauge theories,''
  JHEP {\bf 0012} (2000) 002
  [arXiv:hep-th/0002075].

\bibitem{Minwalla:1999px}
  S.~Minwalla, M.~Van Raamsdonk and N.~Seiberg,
  ``Noncommutative perturbative dynamics,''
  JHEP {\bf 0002} (2000) 020
  [arXiv:hep-th/9912072].


\bibitem{Kontsevich:1997vb}
  M.~Kontsevich,
  ``Deformation quantization of Poisson manifolds, I,''
  Lett.\ Math.\ Phys.\  {\bf 66} (2003) 157
  [arXiv:q-alg/9709040].


\bibitem{klammer} in preparation

\bibitem{Banerjee:2004rs}
  R.~Banerjee and H.~S.~Yang,
  ``Exact Seiberg-Witten map, induced gravity and topological invariants in
  noncommutative field theories,''
  Nucl.\ Phys.\  B {\bf 708} (2005) 434


\bibitem{Madore:2000aq}
  M.~Buric, J.~Madore and G.~Zoupanos,
  ``The Energy-momentum of a Poisson structure,''
  arXiv:0709.3159 [hep-th];
 J.~Madore and J.~Mourad, 
  ``Quantum space-time and classical gravity,''
  J.\ Math.\ Phys.\  {\bf 39} (1998) 423
  [arXiv:gr-qc/9607060]
  

\bibitem{Gilkey:1995mj}
  P.~B.~Gilkey,
  ``Invariance theory, the heat equation and the 
  Atiyah-Singer index theorem,''
  Wilmington, Publish or Perish, 1984


\bibitem{Seiberg:1999vs}
  N.~Seiberg and E.~Witten,
  ``String theory and noncommutative geometry,''
  JHEP {\bf 9909} (1999) 032
  [arXiv:hep-th/9908142].


\bibitem{Cornalba:1999ah}
  L.~Cornalba,
  ``D-brane physics and noncommutative Yang-Mills theory,''
  Adv.\ Theor.\ Math.\ Phys.\  {\bf 4} (2000) 271
  [arXiv:hep-th/9909081].

\bibitem{Jurco:2000fb}
  B.~Jurco and P.~Schupp,
  ``Noncommutative Yang-Mills from equivalence of star products,''
  Eur.\ Phys.\ J.\  C {\bf 14} (2000) 367
  [arXiv:hep-th/0001032].


\bibitem{Hayakawa:1999zf}
  M.~Hayakawa,
  ``Perturbative analysis on infrared and ultraviolet aspects of
  noncommutative QED on $\R^4$,''
  arXiv:hep-th/9912167.

\bibitem{Khoze:2000sy}
  V.~V.~Khoze and G.~Travaglini,
  ``Wilsonian effective actions and the IR/UV mixing in noncommutative  gauge
  theories,''
  JHEP {\bf 0101}, 026 (2001)
  [arXiv:hep-th/0011218].

\bibitem{VanRaamsdonk:2001jd}
  M.~Van Raamsdonk,
  ``The meaning of infrared singularities in noncommutative gauge theories,''
  JHEP {\bf 0111} (2001) 006
  [arXiv:hep-th/0110093].

\bibitem{Armoni:2001uw}
  A.~Armoni and E.~Lopez,
  ``UV/IR mixing via closed strings and tachyonic instabilities,''
  Nucl.\ Phys.\  B {\bf 632} (2002) 240
  [arXiv:hep-th/0110113];
  A.~Armoni, E.~Lopez and A.~M.~Uranga,
  ``Closed strings tachyons and non-commutative instabilities,''
  JHEP {\bf 0302} (2003) 020
  [arXiv:hep-th/0301099].
  
\bibitem{Sarkar:2005jw}
  S.~Sarkar and B.~Sathiapalan,
  ``Aspects of open-closed duality in a background B-field,''
  JHEP {\bf 0505} (2005) 062
  [arXiv:hep-th/0503009];
  S.~Sarkar and B.~Sathiapalan,
  ``Aspects of open-closed duality in a background B-field. II,''
  JHEP {\bf 0511} (2005) 002
  [arXiv:hep-th/0508004];
  S.~Sarkar,
  ``Closed string exchanges on C**2/Z(2) in a background B-field,''
  JHEP {\bf 0605} (2006) 020
  [arXiv:hep-th/0602147];
  S.~Sarkar,
  ``UV / IR mixing in noncommutative field theories and open closed string
  duality,''
  Int.\ J.\ Mod.\ Phys.\  A {\bf 21} (2006) 4763
  [arXiv:hep-th/0606002].


\bibitem{Ambjorn:1999ts}
  J.~Ambjorn, Y.~M.~Makeenko, J.~Nishimura and R.~J.~Szabo,
  ``Finite N matrix models of noncommutative gauge theory,''
  JHEP {\bf 9911} (1999) 029
  [arXiv:hep-th/9911041].


\bibitem{Grosse:2004wm}
  H.~Grosse and H.~Steinacker,
  ``Finite gauge theory on fuzzy $\C P^2$,''
  Nucl.\ Phys.\  B {\bf 707} (2005) 145
  [arXiv:hep-th/0407089];
  W.~Behr, F.~Meyer and H.~Steinacker,
  ``Gauge theory on fuzzy $S^2 \times S^2$ and regularization 
  on noncommutative $\R^4$,''
  JHEP {\bf 0507} (2005) 040
  [arXiv:hep-th/0503041].

\bibitem{Gubser:2000cd}
  S.~S.~Gubser and S.~L.~Sondhi,
  ``Phase structure of non-commutative scalar field theories,''
  Nucl.\ Phys.\  B {\bf 605} (2001) 395
  [arXiv:hep-th/0006119]

\bibitem{Bietenholz:2006cz}
  W.~Bietenholz, J.~Nishimura, Y.~Susaki and J.~Volkholz,
  ``A non-perturbative study of 4d U(1) non-commutative gauge theory: The fate
  of one-loop instability,''
  JHEP {\bf 0610} (2006) 042
  [arXiv:hep-th/0608072].


\bibitem{weyl} 
  H. Weyl, "Das asymptotische Verteilungsgesetz der Eigenwerte linearer
  partieller Differentialoperatoren." 
  {\em Math. Ann.} {\bf 71} (1911), 441

\bibitem{Ishibashi:1996xs}
  N.~Ishibashi, H.~Kawai, Y.~Kitazawa and A.~Tsuchiya,
  ``A large-N reduced model as superstring,''
  Nucl.\ Phys.\  B {\bf 498} (1997) 467
  [arXiv:hep-th/9612115].

\bibitem{Aoki:1999vr}
  H.~Aoki, N.~Ishibashi, S.~Iso, H.~Kawai, Y.~Kitazawa and T.~Tada,
  ``Noncommutative Yang-Mills in IIB matrix model,''
  Nucl.\ Phys.\  B {\bf 565} (2000) 176
  [arXiv:hep-th/9908141].

\bibitem{Ishibashi:2000hh}
  N.~Ishibashi, S.~Iso, H.~Kawai and Y.~Kitazawa,
  ``String scale in noncommutative Yang-Mills,''
  Nucl.\ Phys.\  B {\bf 583} (2000) 159
  [arXiv:hep-th/0004038].

\bibitem{Kitazawa:2005ih}
  Y.~Kitazawa and S.~Nagaoka,
  ``Graviton propagators on fuzzy G/H,''
  JHEP {\bf 0602} (2006) 001
  [arXiv:hep-th/0512204].
  Y.~Kitazawa and S.~Nagaoka,
  ``Graviton propagators in supergravity and noncommutative gauge theory,''
  Phys.\ Rev.\  D {\bf 75} (2007) 046007
  [arXiv:hep-th/0611056].

\bibitem{Bern:2002kj}
  Z.~Bern,
  ``Perturbative quantum gravity and its relation to gauge theory,''
  Living Rev.\ Rel.\  {\bf 5} (2002) 5
  [arXiv:gr-qc/0206071];
  H.~Kawai, D.~C.~Lewellen and S.~H.~H.~Tye,
  ``A Relation Between Tree Amplitudes Of Closed And Open Strings,''
  Nucl.\ Phys.\  B {\bf 269} (1986) 1.


\bibitem{Azuma:2002af}
  T.~Azuma and H.~Kawai,
  ``Matrix model with manifest general coordinate invariance,''
  Phys.\ Lett.\  B {\bf 538} (2002) 393
  [arXiv:hep-th/0204078];
  M.~Hanada, H.~Kawai and Y.~Kimura,
  ``Describing curved spaces by matrices,''
  Prog.\ Theor.\ Phys.\  {\bf 114} (2006) 1295
  [arXiv:hep-th/0508211];
  K.~Furuta, M.~Hanada, H.~Kawai and Y.~Kimura,
  ``Field equations of massless fields in the new interpretation of the matrix
  model,''
  Nucl.\ Phys.\  B {\bf 767} (2007) 82
  [arXiv:hep-th/0611093].

\bibitem{Grosse:2007dm}
   H.~Grosse and M.~Wohlgenannt,
   ``Induced Gauge Theory on a Noncommutative Space,''
   Eur.\ Phys.\ J.\  C {\bf 52} (2007) 435
   [arXiv:hep-th/0703169].



\end{thebibliography}
\end{document}